\documentclass{aa}

\usepackage{graphicx}
\usepackage{txfonts}
\usepackage{pifont}

\newcommand{\h}{^{\text{h}}}
\newcommand{\m}{^{\text{m}}}

\newcommand{\Msun}{\ensuremath{\mathrm{M}_\sun}}

\newcommand{\kms}{\ensuremath{\mathrm{km~s^{-1}}}}

\newcommand{\jy}{\ensuremath{\mathrm{Jy}}}
\newcommand{\mjy}{\ensuremath{\mathrm{m\jy{}}}}
\newcommand{\jyb}{\ensuremath{\mathrm{\jy{}~beam^{-1}}}}
\newcommand{\mjyb}{\ensuremath{\mathrm{m\jyb{}}}}

\newcommand{\ceo}{$\mathrm{C^{18}O}$}

\newcommand{\ceoto}{\ceo{}~($2-1$)}
\newcommand{\tco}{$\mathrm{^{13}CO}$}
\newcommand{\tcoto}{\tco{}~($2-1$)}
\newcommand{\sofs}{\rev{SO~($5_{6}-4_{5}$)}}

\newcommand{\vthau}{\ensuremath{v_\mathrm{200au}}}
\newcommand{\Mstar}{\ensuremath{M_\star}}
\newcommand{\Msinisqr}{\ensuremath{\Mstar \, \left( \sin{i} \right)^{2}}}
\newcommand{\redchi}{\ensuremath{\chi_\nu^2}}
\newcommand{\vlsr}{\ensuremath{v_\mathrm{LSR}}}
\newcommand{\vsys}{\ensuremath{v_\mathrm{sys}}}

\newcommand{\cmark}{\ding{51}}
\newcommand{\xmark}{\ding{55}}

\newcommand\rev[1]{#1}

\begin{document}

\title{Searching for kinematic evidence of Keplerian disks around Class~0 protostars with CALYPSO}

\author{S.~Maret\inst{\ref{ipag}} \and
  A.~J.~Maury\inst{\ref{cea}} \and
  A.~Belloche\inst{\ref{mpifr}} \and
  M.~Gaudel\inst{\ref{cea}} \and
  Ph.~André\inst{\ref{cea}} \and
  S.~Cabrit\inst{\ref{lerma}} \and
  C.~Codella\inst{\ref{ipag},\ref{arcetri}} \and
  C.~Lefèvre\inst{\ref{iram}} \and
  L.~Podio\inst{\ref{arcetri}} \and
  S.~Anderl\inst{\ref{ipag}} \and
  F.~Gueth\inst{\ref{iram}} \and
  P.~Hennebelle\inst{\ref{cea}}
}
\institute{Univ. Grenoble Alpes, CNRS, IPAG, 38000 Grenoble,
  France\label{ipag}
  \and
  AIM, CEA, CNRS, Université Paris-Saclay, Université Paris Diderot,
  Sorbonne Paris Cité, 91191 Gif-sur-Yvette, France\label{cea}
  \and
  Max-Planck-Institut für Radioastronomie, Auf dem Hügel 69, 53121
  Bonn, Germany\label{mpifr}
  \and
  LERMA, Observatoire de Paris, PSL Research University, CNRS,
  Sorbonne Université, UPMC Université Paris 06, 75014 Paris,
  France\label{lerma}
  \and
  INAF - Osservatorio Astrofisico di Arcetri,
  Largo E. Fermi 5, 50125 Firenze, Italy\label{arcetri}
  \and
  Institut de Radioastronomie Millimétrique (IRAM), 38406
  Saint-Martin-d'Hères, France\label{iram}
}
\authorrunning{Maret et al.}

\date{Received September 27, 2019; accepted January 6, 2020}

\abstract{The formation of protoplanetary disks is not well
  understood. To understand how and when these disks are formed, it is
  crucial to characterize the kinematics of the youngest protostars at
  a high angular resolution.  Here we study a sample of 16 Class 0
  protostars to measure their rotation profile at scales from 50 to
  500~au and search for Keplerian rotation. We used
  high-angular-resolution line observations obtained with the Plateau
  de Bure Interferometer as part of the CALYPSO large program. From
  \rev{\tco{}~($J=2-1$), \ceo{}~($J=2-1$) and
  SO~($N_{j}=5_{6}-4_{5}$)} moment maps, we find that seven sources
  show rotation about the jet axis at a few hundred au scales:
  SerpS-MM18, L1448-C, L1448-NB, L1527, NGC1333- IRAS2A,
  NGC1333-IRAS4B, and SVS13-B. We analyzed the kinematics of these
  sources in the $uv$ plane to derive the rotation profiles down to
  50~au scales. We find evidence for Keplerian rotation in only two
  sources, L1527 and L1448-C. Overall, this suggests that Keplerian
  disks larger than 50~au are uncommon around Class 0
  protostars. However, in some of the sources, the line emission could
  be optically thick and dominated by the envelope emission. \rev{Due
  to the optical thickness of these envelopes,} some of the disks
  could have remained undetected in our observations.
 }


\maketitle

\section{Introduction}
\label{sec:introduction}

Protoplanetary disks are the birth site of planetary
systems. Determining how and when these disks are formed is,
therefore, important for understanding the formation of planetary
systems. Protoplanetary disks are ubiquitous around Class II young
stellar objects. Continuum observations at millimeter to infrared
wavelengths indicate that their radii comprise between a few tens and
a few hundreds of au \citep[see][for a review]{Williams11}, but recent
ALMA surveys also find disks with smaller radii
\citep[e.g.\rev{,}][]{2019ApJ...882...49L}. Millimeter line
observations show that the gas around these disks is in Keplerian
motion \citep[e.g.\rev{,}][]{Simon00} and, therefore, these disks are
rotationally supported. Disks are also common around Class I
protostars
\citep[e.g.\rev{,}][]{Brinch07,Takakuwa12,2016ApJ...826..213L}. However,
disks around the youngest Class 0 protostars are more difficult to
observe. Because these objects are the youngest accreting protostars,
most of their mass is still in the form of an envelope
\citep{1993ApJ...406..122A,2000prpl.conf...59A,2014prpl.conf..195D}\rev{,}
which makes the detection of a disk challenging. Yet the observation
and the characterization of disks around young embedded protostars are
key to understanding the formation of disks.

In the absence of magnetic fields, disks are expected to form and to
grow quickly as a consequence of angular momentum conservation during
the rotating collapse of the protostellar envelope
\citep{1984ApJ...286..529T,1999ApJ...525..330Y}. However, magnetic
fields can redistribute angular momentum from small to large scales
efficiently (through so-called magnetic braking). Early
magneto-hydrodynamical (MHD) simulations (in the ideal limit) have
shown that the disk formation is hampered, even for modest values of
the magnetic flux
\citep{2003ApJ...599..363A,2006ApJ...647..374G,Hennebelle08a,2008ApJ...681.1356M}. More
recent simulations that consider non-ideal MHD effects, radiative
transfer, turbulence, and different initial conditions (such as the
misalignment between the magnetic field and the rotation axis) now
predict the formation of small (with a radius \rev{of} a few tens of
au) disks during the Class 0 phase
\citep[e.g.\rev{,}][]{2016ApJ...830L...8H}.

Observational evidence for disks around Class 0 protostars is still
scarce. \citet{Jorgensen09} observed a sub-millimeter continuum
emission excess, mostly unresolved within the \rev{Submillimeter
  Array} (SMA) beam, in a sample of Class 0 protostars that they
attributed to disk emission. \citet{Yen15a} used SMA observations to
derive the rotation of a sample of 17 protostars on 1000 au
scales. Assuming that the specific angular momentum was conserved
during the collapse, they derived centrifugal radii comprised between
5 and 500 au. However, the spatial resolution of their observations is
not sufficient for the detection of Keplerian rotation within the
centrifugal radius. So far, direct evidence of Keplerian motions has
only been found in a few Class 0 protostars and that is thanks to high
resolution (sub-)millimeter interferometric observations: L1527, a
border-line Class 0 object \citep[][disk radius of
74~au]{Tobin12b,Ohashi14,2017ApJ...849...56A}, VLA1623, the
prototypical Class 0 protostar \citep[][disk radius of
150~au]{Murillo13}, HH212 \citep[][disk radius of
90-120~au]{Codella14,Lee14}, L1448-IRS3 \citep[disk radius of
$\sim$400~au]{2016Natur.538..483T}, Lupus~3~MMS \citep[][disk radius
of 100~au]{2017ApJ...834..178Y}, and IRAS~16253-2429 \citep[][disk
radius of 8-32~au]{2019ApJ...871..100H}. On the other hand, no disks
have been detected at scales larger than 100~au in NGC1333-IRAS2A
\citep{Maret14} and 10~au in B335 \citep{Yen15b}. Therefore, it is
still unclear if Keplerian disks are common around Class 0 protostars.

\begin{table*}
  \begin{center}
    \caption{Source sample.\label{tab:sources}}
    \begin{tabular}{lllllllllll}
      \hline
      \hline
      Source &  Peak & \multicolumn{2}{c}{Position (J2000)\tablefootmark{a}} & dist. & $L_\mathrm{int}$\tablefootmark{b} & $M_\mathrm{env}$\tablefootmark{c}
      & $S_\mathrm{1.4 mm}$\tablefootmark{d} & jet PA\tablefootmark{e} & Ref.\tablefootmark{f} \\
      \cline{3-4}
             &  & R.A. & Dec. & (pc) & (L$_\odot$) & (M$_\odot$) & (\mjy{} & ($\degr$)      &  \\
             &  &      &      &      &             &             &   beam$^{-1}$ )      &  & \\
      \hline
      L1448-2A       & 2A      & 03$\h$25$\m$22$\fs$405 & 30$\degr$45$\arcmin$13$\farcs$26  & 293                  & 4.7   & 1.9  & 23  & -63\tablefootmark{g} & 1, 2, 3    \\
                     & 2Ab     & 03$\h$25$\m$22$\fs$355 & 30$\degr$45$\arcmin$13$\farcs$16  & 293                  & ...   & ...  & 11  & ...                  & 1, 2, ...  \\
      L1448-NB       & NB1     & 03$\h$25$\m$36$\fs$378 & 30$\degr$45$\arcmin$14$\farcs$77  & 293                  & 3.9   & 4.8  & 146 & -80                  & 4, 2, 5    \\
                     & NB2     & 03$\h$25$\m$36$\fs$315 & 30$\degr$45$\arcmin$15$\farcs$15  & 293                  & ...   & ...  & 69  & ...                  & 4, 2, ...  \\
      L1448-C        &         & 03$\h$25$\m$38$\fs$875 & 30$\degr$44$\arcmin$05$\farcs$33  & 293                  & 10.9  & 2.0  & 123 & -17                  & 6, 2, 5    \\
      NGC1333-IRAS2A &         & 03$\h$28$\m$55$\fs$570 & 31$\degr$14$\arcmin$37$\farcs$07  & 293                  & 47    & 7.9  & 132 & -155                 & 7, 2, 8    \\
      SVS13B         &         & 03$\h$29$\m$03$\fs$078 & 31$\degr$15$\arcmin$51$\farcs$74  & 293                  & 3.1   & 2.8  & 127 & 167                  & 9, 2, 10   \\
      NGC1333-IRAS4A & 4A1     & 03$\h$29$\m$10$\fs$537 & 31$\degr$13$\arcmin$30$\farcs$98  & 293                  & 4.7   & 12.3 & 481 & 180                  & 7, 2, 5    \\
                     & 4A2     & 03$\h$29$\m$10$\fs$432 & 31$\degr$13$\arcmin$32$\farcs$12  & 293                  & ...   & ...  & 186 & -178                 & 7, 2, ...  \\
      NGC1333-IRAS4B &         & 03$\h$29$\m$12$\fs$016 & 31$\degr$13$\arcmin$08$\farcs$02  & 293                  & 2.3   & 4.7  & 278 & 167                  & 7, 2, 5    \\
      IRAM04191      &         & 04$\h$21$\m$56$\fs$899 & 15$\degr$29$\arcmin$46$\farcs$11  & 140                  & 0.05  & 0.5  & 4.7 & -160                 & 11, 12, 13 \\
      L1521F         &         & 04$\h$28$\m$38$\fs$941 & 26$\degr$51$\arcmin$35$\farcs$14  & 140                  & 0.035 & 0.7  & 1.6 & -120                 & 14, 12, 15 \\
      L1527          &         & 04$\h$39$\m$53$\fs$875 & 26$\degr$03$\arcmin$09$\farcs$66  & 140                  & 0.9   & 1.2  & 129 & 90                   & 16, 12, 17 \\
      SerpM-S68N     &         & 18$\h$29$\m$48$\fs$091 & 01$\degr$16$\arcmin$43$\farcs$41  & 436                  & 11    & 11   & 35  & -45                  & 18, 19, 20 \\
      SerpM-SMM4     & \rev{a} & 18$\h$29$\m$56$\fs$716 & 01$\degr$13$\arcmin$15$\farcs$65  & 436                  & 2     & 8    & 184 & 10                   & 18, 19, 20 \\
      SerpS-MM18     & \rev{a} & 18$\h$30$\m$04$\fs$118 & -02$\degr$03$\arcmin$02$\farcs$55 & 350                  & 29    & 5    & 148 & -172                 & 21, 22, 21 \\
      SerpS-MM22     &         & 18$\h$30$\m$12$\fs$310 & -02$\degr$06$\arcmin$53$\farcs$56 & 350                  & 0.4   & 0.9  & 20  & -130                 & 21, 22, 21 \\
      L1157          &         & 20$\h$39$\m$06$\fs$269 & 68$\degr$02$\arcmin$15$\farcs$70  & 352                  & 4.0   & 3.0  & 117 & 163                  & 23, 24, 17 \\
      GF9-2          &         & 20$\h$51$\m$29$\fs$823 & 60$\degr$18$\arcmin$38$\farcs$44  & 200\tablefootmark{h} & 0.3   & 0.5  & 9.9 & 0                    & 25, 26, 26 \\
      \hline
    \end{tabular}  
    \tablefoot{
      \tablefoottext{a}{Equatorial coordinates of the continuum peak measured by \citet{2019A&A...621A..76M}.}
      \tablefoottext{b}{Internal luminosity, estimated from the Herschel
        Gould Belt Survey (HGBS, see
        \citealt{2010A&A...518L.102A} and Ladjelate et al. in
        prep.). }
      \tablefoottext{c}{Envelope mass, corrected for the assumed distance.}
      \tablefoottext{d}{Peak flux density at 1.4~mm measured by \citet{2019A&A...621A..76M}.}
      \tablefoottext{e}{PA of the jet blue lobe. \rev{We adopt the
          values measured by Podio et al. (in prep.) from CALYPSO
          observations of the SiO line emission. These values are in
          agreement with those estimated from previous studies,
          e.g.\rev{,} \citet[NGC1333-IRAS2A;][]{Codella14},
          \citet[NGC1333-IRASAA;][]{2015A&A...584A.126S},
          \citet[L1157;][]{2016A&A...593L...4P} and
          \citet[L1521F;][]{2014ApJ...789L...4T,2016ApJ...826...26T}.}}
      \tablefoottext{f}{References for the protostar discovery, distance and envelope mass.}
      \tablefoottext{g}{Jet is asymmetric: the red-shifted lobe
        PA is 140\degr.}
      \tablefoottext{h}{The distance of GF9-2 is very uncertain. Here
        we adopt a distance of 200~pc \citep{1997A&A...320..287W}, but
        another study suggests a distance of 474~pc (Zucker,
        priv. comm.).}
    }
    \tablebib{
      (1) \cite{1999ApJ...515..696O}; (2) \cite{2018ApJ...865...73O};
      (3) \cite{Enoch09}; (4) \cite{1990ApJ...365L..85C}; (5)
      \cite{Sadavoy14}; (6) \cite{1989ApJ...341..208A}; (7)
      \cite{1987MNRAS.226..461J}; (8) \cite{2013A&A...552A.141K}; (9)
      \cite{1987ApJ...320..356G}; (10) \cite{1997A&A...325..542C};
      (11) \cite{1999ApJ...513L..57A}; (12) \cite{Loinard07}; (13)
      \cite{2000prpl.conf...59A}; (14) \cite{1994Natur.368..719M};
      (15) \cite{2016ApJ...826...26T}; (16)
      \cite{1991ApJ...382..555L}; (17) \cite{2001A&A...365..440M};
      (18) \cite{1993A&A...275..195C}; (19)
      \cite{2017ApJ...834..143O}; (20) \cite{2004A&A...421..623K};
      (21) \cite{2011A&A...535A..77M}; (22) Palmeirim et al. (in prep.); (23)
      \cite{1992ApJ...392L..83U}; (24) \citet{2019ApJ...879..125Z};
      (25) \cite{1979ApJS...41...87S}; (26)
      \cite{1997A&A...320..287W}.
      }
  \end{center}
\end{table*}

In this paper, we present high-angular millimeter line interferometric
observations of a sample of 16 Class 0 protostars obtained with the
IRAM Plateau de Bure interferometer (hereafter PdBI) as part of the
CALYPSO (Continuum and Line from Young Protostellar Objects)
survey. CALYPSO is an IRAM Large Program (P.I. Philippe André) that
consists of line and continuum observations towards a large sample of
nearby ($d < 436$~pc) Class 0 protostars that aims to understand the
angular momentum problem\footnote{See
  \url{http://irfu.cea.fr/Projets/Calypso} for more details.}. Here we
use line observations to measure the rotation at scales between 50 and
500~au and to determine whether Keplerian disks are present in these
objects. In a recent paper, \citet{2019A&A...621A..76M} study the
continuum emission in the same sample to determine if disk-like
structures are present. A companion paper focuses on the rotation of
the envelopes on scales up to 5000~au (\citeauthor{Gaudel2020} subm.).

\section{Observations}
\label{sec:observations}

Observations were carried out with the PdBI between September 2010 and
March 2013 as part of CALYPSO\footnote{CALYPSO data are available from
  \url{http://www.iram.fr/ILPA/LP010/}.}. We observed a sample of 16
Class 0 protostars located in the Taurus, Perseus, Serpens, and Aquila
molecular cloud complexes at distances between 140 and
436~pc. Table~\ref{tab:sources} gives the continuum peak positions,
distances, internal luminosities, envelope masses, and flux densities
at 1.4~mm of the sources. Our sample covers a wide range of internal
luminosities and envelope masses and should, therefore, be
representative of the whole population of Class 0 protostars.

We used the configurations A and C of the array, which provide
baselines\footnote{These baselines correspond to physical scales
  between 0.4\arcsec{} and 19\arcsec{} at a wavelength of 1.4~mm.}
between 15 m and 760 m. In this study, we focus on the \tcoto{}
(220.398\,684~GHz), \ceoto{} (219.560\,354~GHz) and \sofs{}
(219.949\,442~GHz) line emission. Previous studies have shown that
these lines are good tracers of disks in Class 0 protostars
\citep{Tobin12b,Murillo13,Ohashi14,Sakai14}. These three lines were
observed simultaneously using the 1.3~mm receivers connected to the
narrow-band backend with a bandwidth of 512 channels of 39~kHz
(0.05~\kms{}) each.

The data were calibrated and reduced using the {\tt GILDAS} software
package \citep{Gildas}. For sources with a continuum peak flux density
higher than 80~\mjyb{}, the line observations were self-calibrated
using the continuum visibilities\footnote{We refer the reader to
  \citet{2019A&A...621A..76M} for details on the calibration of the
  continuum observations.}. Other sources were calibrated using strong
quasars only. To improve the signal-to-noise ratio, all line
observations were re-sampled to a spectral resolution of
0.2~\kms{}. Imaging was done using robust weighting \rev{with a robust
  parameter of 1}. Typical synthesized beam sizes are 0.7$\arcsec$
(HPBW), which corresponds to linear scales between 100 and
\rev{300}~au, depending on the source distance. The sensitivity of the
line observations is between 10 and 24~\mjyb{} per 0.2~\kms{} channel
depending on the source. Synthesized beams and sensitivities for each
line are given in Appendix~\ref{sec:synth-beam-sens}.

\section{Results}
\label{sec:results}

\subsection{Overview of the sample}
\label{sec:overview-sample}

\begin{figure*}
  \includegraphics[width=\hsize]{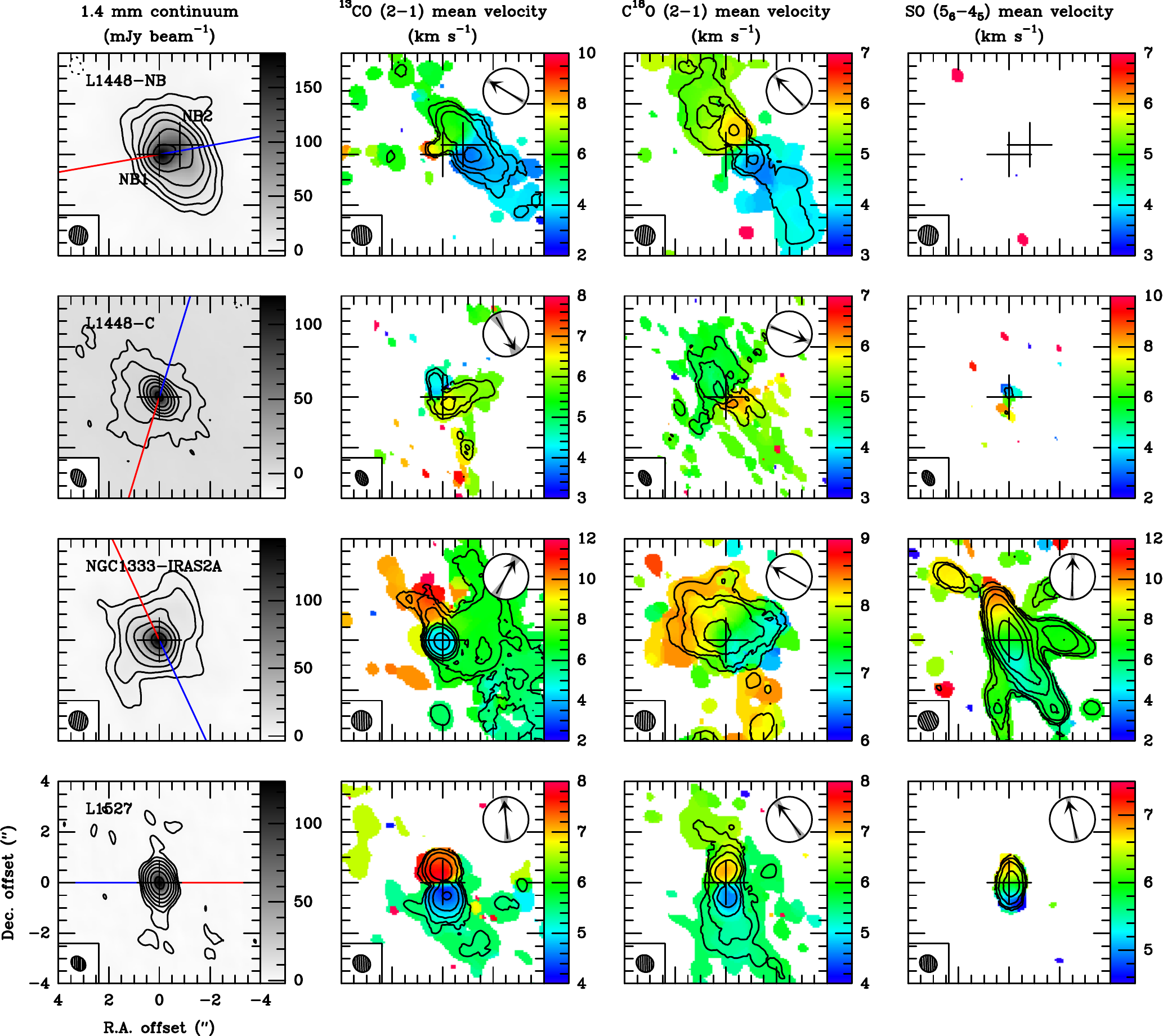}
  \caption{1.4 mm continuum emission\rev{, integrated intensities} and
    mean velocities of the \tcoto{}, \ceoto{} and \sofs{} lines in
    L1448-NB, L1448-C, NGC1333-IRAS2A, and L1527. Left panels show the
    1.4 mm continuum emission. The other panels show the line mean
    velocities (first-order moment; background image), together with
    the line \rev{integrated} intensities (zeroth-order moment; black
    contours). Contours are drawn at -3$\sigma$ (dotted lines)
    3$\sigma$, 6$\sigma$, 12$\sigma$ and so on (solid lines). In each
    panel, the black crosses show the position of the main continuum
    peak(s). The blue and red lines in the left panels indicate the
    direction of the blue- and red-shifted lobes of the jet(s),
    respectively.  The arrows in other panels indicate the direction
    of the velocity gradient, as determined from a linear fit of the
    mean velocity within 2\arcsec{} of the continuum peak
    position. The gray sectors around each arrow show the 1$\sigma$
    uncertainty on the direction of the velocity gradient. The ellipse
    in each panel shows the synthesized beam. All coordinates are
    relative to the positions of the brightest continuum peak of each
    source (see Table~\ref{tab:sources}).\label{momentmaps-1}}
\end{figure*}

\begin{table*}
  \begin{center}
    \caption{Results of linear fit of the mean velocity for each line. \label{tab:moment1}}
    \begin{tabular}{llllllll}
      \hline
      \hline
      Source         & Line     & $G$                     & $v_{0}$         & $\theta$      & $\Delta \theta$ \\ 
                     &          & (km s$^{-1}$ pc$^{-1}$) & (km s$^{-1}$)   & ($\degr{}$)   & ($\degr{}$)     \\
      \hline
      L1448-2A       & \tcoto{} & 465 $\pm$ 58   & 4.18 $\pm$ 0.07  & 115 $\pm$ 7   & 65  \\
                     & \ceoto{} & 210 $\pm$ 15   & 4.23 $\pm$ 0.02  & 113 $\pm$ 5   & 63  \\
                     & \sofs{}  & ...            & ...              & ...           & ... \\
      \hline
      L1448-NB       & \tcoto{} & 863 $\pm$ 111  & 4.97 $\pm$ 0.15  & 59 $\pm$ 7    & 49  \\
                     & \ceoto{} & 453 $\pm$ 42   & 4.85 $\pm$ 0.06  & 44 $\pm$ 5    & 34  \\
                     & \sofs{}  & ...            & ...              & ...           & ... \\
      \hline
      L1448-C        & \tcoto{} & 526 $\pm$ 133  & 4.97 $\pm$ 0.11  & -150 $\pm$ 15 & 43  \\
                     & \ceoto{} & 279 $\pm$ 35   & 5.09 $\pm$ 0.04  & -111 $\pm$ 6  & 4   \\
                     & \sofs{}  & ...            & ...              & ...           & ... \\
      \hline
      NGC1333-IRAS2A & \tcoto{} & 968 $\pm$ 205  & 5.07 $\pm$ 0.13  & -29 $\pm$ 12  & 36  \\
                     & \ceoto{} & 358 $\pm$ 23   & 7.52 $\pm$ 0.03  & 60 $\pm$ 4    & 55  \\
                     & \sofs{}  & 1033 $\pm$ 70  & 6.86 $\pm$ 0.06  & -1 $\pm$ 6    & 64  \\
      \hline
      SVS13B         & \tcoto{} & ...            & ...              & ...           & ... \\
                     & \ceoto{} & 161 $\pm$ 30   & 8.48 $\pm$ 0.02  & 92 $\pm$ 9    & 15  \\
                     & \sofs{}  & ...            & ...              & ...           & ... \\
      \hline
      NGC1333-IRAS4A & \tcoto{} & 769 $\pm$ 125  & 6.21 $\pm$ 0.13  & 17 $\pm$ 6    & 73  \\
                     & \ceoto{} & 116 $\pm$ 38   & 6.26 $\pm$ 0.03  & -36 $\pm$ 19  & 54  \\
                     & \sofs{}  & 373 $\pm$ 126  & 0.67 $\pm$ 0.18  & -145 $\pm$ 18 & 55  \\
      \hline
      NGC1333-IRAS4B & \tcoto{} & 206 $\pm$ 27   & 6.28 $\pm$ 0.05  & -7 $\pm$ 19   & 84  \\
                     & \ceoto{} & 70 $\pm$ 10    & 6.81 $\pm$ 0.01  & -79 $\pm$ 6   & 24  \\
                     & \sofs{}  & 4093 $\pm$ 447 & 6.75 $\pm$ 0.26  & 74 $\pm$ 6    & 3   \\
      \hline
      IRAM04191      & \tcoto{} & 182 $\pm$ 153  & 7.74 $\pm$ 0.05  & -119 $\pm$ 49 & 49  \\
                     & \ceoto{} & ...            & ...              & ...           & ... \\
                     & \sofs{}  & ...            & ...              & ...           & ... \\
      \hline
      L1527          & \tcoto{} & 1670 $\pm$ 244 & 5.58 $\pm$ 0.13  & 5 $\pm$ 10    & 5   \\
                     & \ceoto{} & 671 $\pm$ 98   & 5.71 $\pm$ 0.06  & 36 $\pm$ 10   & 36  \\
                     & \sofs{}  & 1816 $\pm$ 152 & 6.05 $\pm$ 0.04  & 12 $\pm$ 7    & 12  \\
      \hline
      SerpM-S68N     & \tcoto{} & 674 $\pm$ 79   & 8.69 $\pm$ 0.19  & 132 $\pm$ 7   & 87  \\
                     & \ceoto{} & 27 $\pm$ 52    & 9.19 $\pm$ 0.05  & 174 $\pm$ 90  & 51  \\
                     & \sofs{}  & ...            & ...              & ...           & ... \\
      \hline
      SerpS-MM18     & \tcoto{} & 675 $\pm$ 194  & 4.50 $\pm$ 0.25  & -159 $\pm$ 26 & 77  \\
                     & \ceoto{} & 362 $\pm$ 19   & 8.06 $\pm$ 0.03  & -42 $\pm$ 3   & 40  \\
                     & \sofs{}  & 2586 $\pm$ 255 & 4.39 $\pm$ 0.34  & -87 $\pm$ 3   & 5   \\
      \hline
      SerpS-MM22     & \tcoto{} & 97 $\pm$ 97    & 5.78 $\pm$ 0.14  & 35 $\pm$ 61   & 75  \\
                     & \ceoto{} & 56 $\pm$ 33    & 6.23 $\pm$ 0.05  & -95 $\pm$ 34  & 55  \\
                     & \sofs{}  & ...            & ...              & ...           & ... \\
      \hline
      L1157          & \tcoto{} & 205 $\pm$ 40   & 2.27 $\pm$ 0.05  & -25 $\pm$ 15  & 82  \\
                     & \ceoto{} & 59 $\pm$ 13    & 2.62 $\pm$ 0.02  & -45 $\pm$ 13  & 62  \\
                     & \sofs{}  & ...            & ...              & ...           & ... \\
      \hline
      GF9-2          & \tcoto{} & 550 $\pm$ 53   & -2.98 $\pm$ 0.04 & -179 $\pm$ 5  & 89  \\
                     & \ceoto{} & ...            & ...              & ...           & ... \\
                     & \sofs{}  & ...            & ...              & ...           & ... \\
      \hline
    \end{tabular}
    \tablefoot{$G$, $v_{0}$ and $\theta$ are the gradient amplitude,
      mean velocity on the source continuum peak position, and
      gradient position angle (from the north to the east),
      respectively.  $\Delta \theta$ is the difference between
      $\theta$ and the \rev{direction} perpendicular to the
      jet. Missing values correspond to lines with insufficient
      signal-to-noise ratio to fit a gradient. L1521F and SerpM-SMM4
      are not included in this Table, because no velocity gradient
      could be fitted for any of the observed lines. }
  \end{center}
\end{table*}

\begin{figure*}
  \includegraphics[width=\hsize]{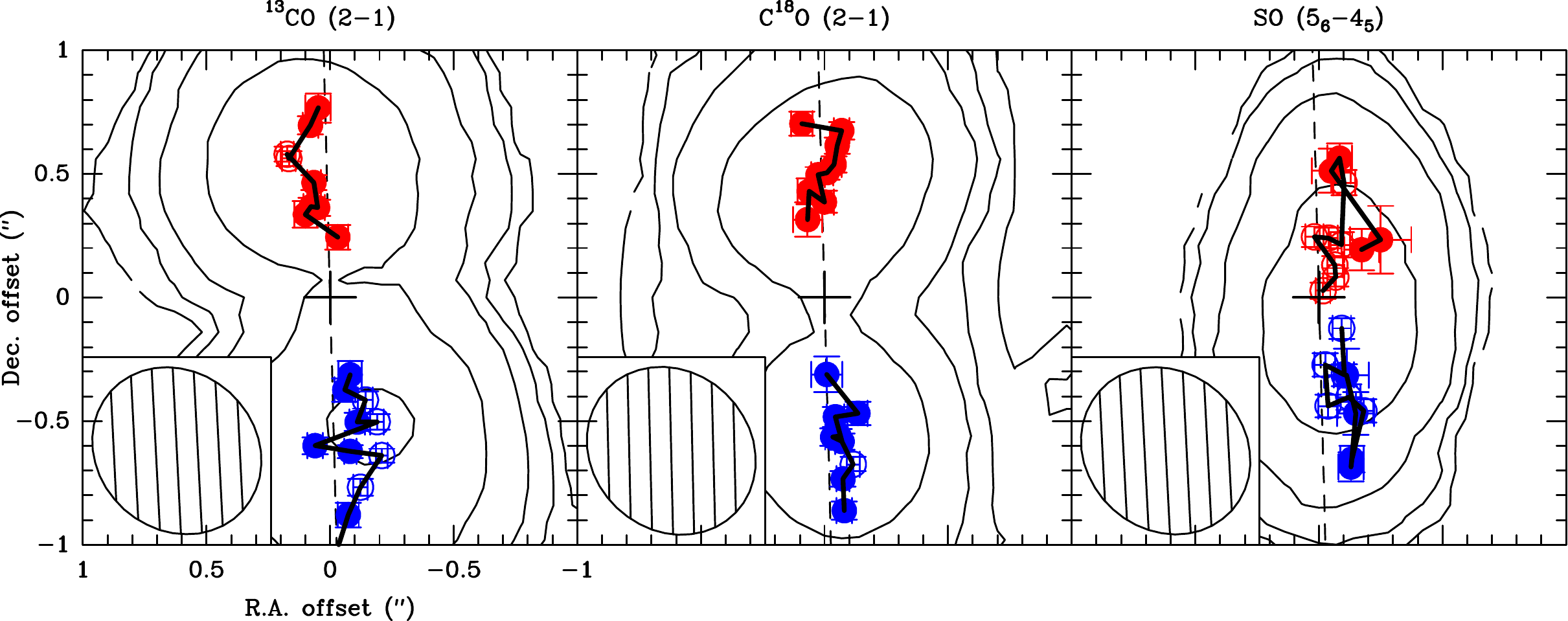}
  \caption{Results of the fits in the $uv$ plane for each spectral
    channel (circles with error bars) superimposed on zeroth-order
    moment map (black contours) in L1527 for the \tcoto{} (left
    panel), \ceoto{} (middle panel), and \sofs{} (right panel)
    lines. Contours are drawn at 3$\sigma$, 6$\sigma$, 12$\sigma,$ and
    so on. Blue and red points show the centroid positions of the
    blue-shifted and red-shifted channels, respectively. Only filled
    points are considered when constructing the velocity curves; open
    points are ignored.  The thick black lines connect the centroids
    that correspond to adjacent velocity channels. Black crosses show
    the position of the continuum peak, while the dashed line shows
    the assumed disk major axis. \label{fig:uvfit-l1527}}
\end{figure*}

\begin{figure*}
  \includegraphics[width=\hsize]{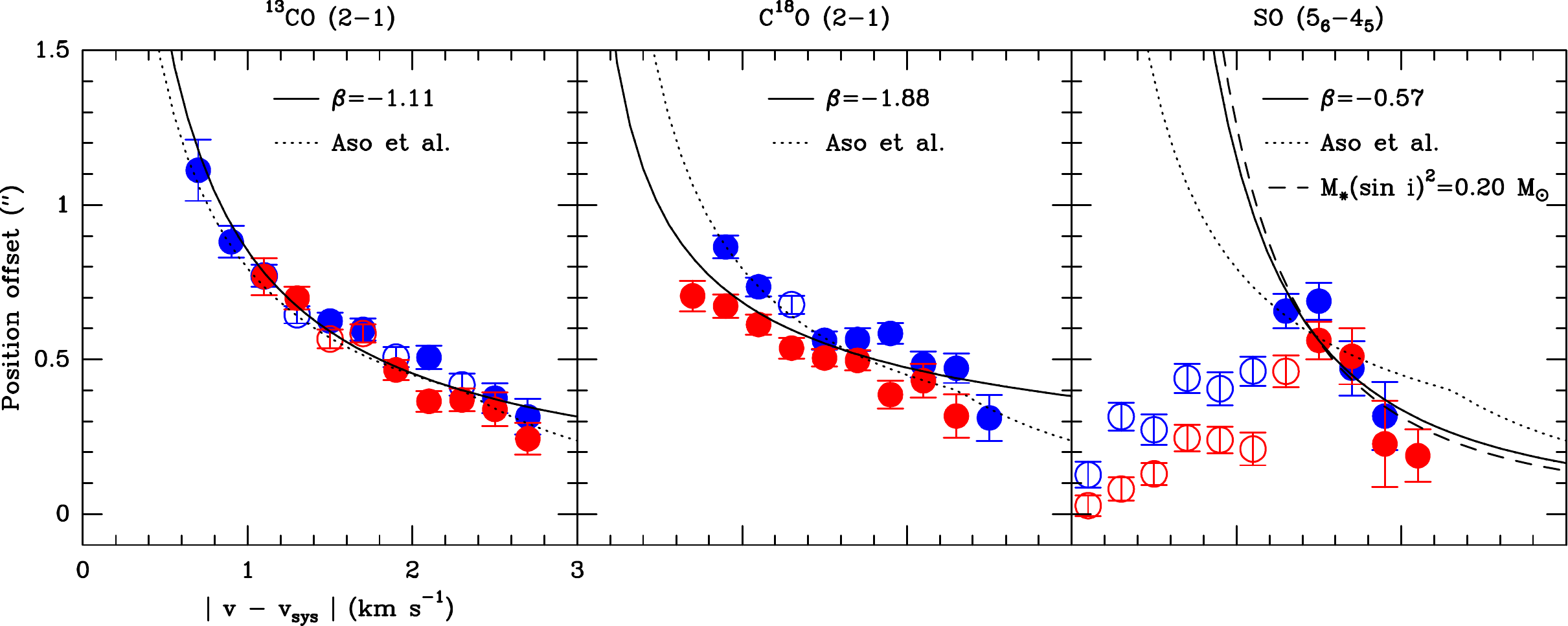}
  \caption{Position offset derived in the $uv$ plane along the disk
    major axis as a function of the velocity channel in L1527 for
    \tcoto{} (left panel) \ceoto{} (middle panel) and \sofs{} (right
    panel).  In each panel the blue and red points with the error bars
    show the observations, the solid line shows the result of a
    power-law fit. The dashed line in the right panel shows the result
    of a fit with a Keplerian law with $\Msinisqr = 0.2$~\Msun{}. Only
    filled points are considered in the power-law and Keplerian
    fits. The dotted lines show the velocity fit from
    \citet{2017ApJ...849...56A}. \label{fig:velocityfit-l1527} }
\end{figure*}

\begin{figure*}
  \includegraphics[width=\hsize]{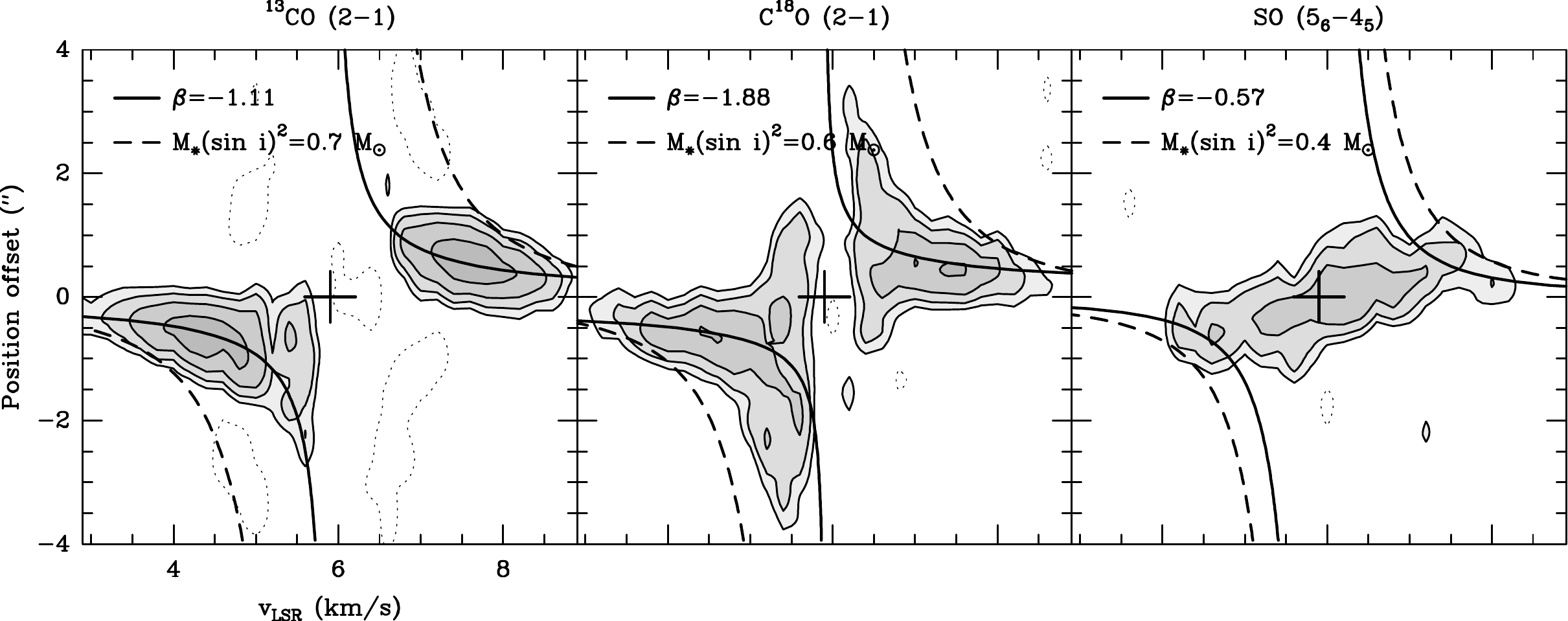}
  \caption{Position velocity cuts in the image plane through the L1527
    continuum peak along the assumed disk major axis for the \tcoto{}
    (left panel), \ceoto{} (middle panel), and \sofs{} (right panel)
    lines. Contours are drawn at -3$\sigma$, 3$\sigma$, 6$\sigma$,
    12$\sigma$ and so on. Black crosses correspond to the continuum
    peak. Black solid curves show the results of the velocity fit in
    the $uv$ plane with a power-law function. Dashed lines show the
    result of the fit of the 6$\sigma$ contours with a Keplerian law
    with $\Msinisqr = 0.7$~\Msun{}, $\Msinisqr = 0.6$~\Msun{}, and
    $\Msinisqr = 0.4$~\Msun{} for the \tcoto{}, \ceoto{}, and \sofs{}
    lines, respectively. \label{fig:pv-l1527}}
\end{figure*}

Figure~\ref{momentmaps-1} shows the zeroth-order moment (integrated
intensity) and first-order moment (mean velocity) maps of the
\tcoto{}, \ceoto{} and \sofs{} lines, together with the 1.4~mm
continuum emission maps from \citet{2019A&A...621A..76M} in L1448-NB,
L1448-C, NGC1333-IRAS2A, and L1527. The same figures for \rev{the}
other sources of the sample are shown in
Appendix~\ref{sec:moment-maps}
(Figs.~\ref{momentmaps-2}-~\ref{momentmaps-4}). The moments are
computed using all spectral channels with emission above 5$\sigma$. In
these figures, we also show for each source, the jet axis as derived
from CALYPSO observations (Podio et al., in prep). As seen in the
zeroth-order moment maps, \tcoto{} emission is detected in all sources
except L1521F, which is, together with IRAM04191, the lowest
luminosity source of our sample. \ceoto{} is also detected in most
sources, except IRAM04191, L1521F, and GF9-2. Finally, we detect
\sofs{} emission \rev{in five sources: NGC1333-IRAS2A, NGC1333-IRAS4A,
  NGC1333-IRAS4B, L1527, and SerpS-MM18.} The morphology of the
emission varies from one source and line to the other and can be
classified into three broad categories: 1. extended emission with
respect to the continuum emission (e.g\rev{.,} the \ceoto{} emission
in NGC1333-IRAS2A) 2. compact emission centered on continuum emission
(e.g.\rev{,} the \tcoto{} emission in L1527) and 3. bi-polar emission
from the outflow (e.g.\rev{,} \sofs{} in NGC1333-IRAS2A). For most
sources, the \tcoto{} and \ceoto{} line emissions fall into the first
two categories. On the other hand, the \sofs{} emission, when
detected, falls into the third category for most sources, with the
notable exception of L1527.

Most lines and sources show clear velocity gradients (as seen from the
first-order moment maps), which could be due to the rotation of
the envelope \rev{or} a disk, or to the \rev{radial motion of} the bipolar
outflow. In order to determine the origin of these gradients, we fit
the first-order moment map of each line with a linear gradient
\citep{Goodman93}:

\begin{equation}
  M_{1} = v_0 + a \, \Delta \alpha + b \, \Delta \delta,
\end{equation}

\noindent
\rev{where $M_{1}$ is the first-order moment, $\Delta \alpha$ and
  $\Delta \delta$ are the R.A. and declination offsets from the
  continuum peak position, and $v_0$, $a$ and $b$ are the fitted
  parameters.} A linear gradient is expected for a spherical envelope
rotating as a solid body. Although solid-body rotation may be a crude
approximation, such a fit provides a rough estimate of the velocity
gradient amplitude and its orientation, and it is useful to determine
the origin of the emission \citep[see\rev{,}
e.g.\rev{,}][]{Yen15a}. To perform these fits, we compute, for each
pixel of \rev{in the image plane}, the first-order moment and its
uncertainty \citep[see ][Eq. 2.3]{Belloche13}, ignoring channels with
emission lower than 5$\sigma$. Because the datacubes produced by the
interferometer are over-sampled spatially, only two points per
synthesized beam are fitted to ensure Nyquist sampling. Since we are
interested in measuring the rotation of the innermost regions of the
protostar (and in turn to determine if a disk is present), we consider
pixels within a radius of 2$\arcsec$ from the continuum peak. This
corresponds to a maximum radius from 300 to 1000~au depending on the
source distance. \rev{The velocity gradient amplitude is given by:}

\begin{equation}
  G = \sqrt{a^{2} + b^{2}} / d,
\end{equation}

\noindent
\rev{where $d$ is the source distance. The position angle of the
  gradient, measured from north to east, is:}

\begin{equation}
  \theta = \mathrm{tan}^{-1} \left( a / b \right).
\end{equation}

Table~\ref{tab:moment1} gives \rev{the values of $G$, $v_0$ and
  $\theta$ that we obtain.} The gradient orientations are also shown
in Fig.~\ref{momentmaps-1} and Figs.~
\ref{momentmaps-2}-\ref{momentmaps-4}. In Table~\ref{tab:moment1} we
also report the difference between the velocity gradient orientation
and the direction perpendicular to the jet, \rev{which} we note as
$\Delta \theta$. A value close to 0\degr{} is expected if the velocity
gradient is due to the rotation of the envelope or a disk, while a
value close to 90\degr{} indicates that the gradient is mostly due to
the outflow, or infall in a flattened geometry. Based on these values,
we selected a sample of disk candidates in which
$\Delta \theta < 45\degr$ for at least one line\footnote{For L1448-2A,
  which has asymmetric jets, we use the direction of the jet lobe that
  gives the lowest $\Delta \theta$ value.}. According to this
criterion, we find seven disk candidates: L1448-NB, L1448-C,
NGC1333-IRAS2A, SVS13B, NGC1333-IRAS4B, \rev{L1527} and
\rev{SerpS-MM18}. In the following, we analyze the line kinematics of
these sources to derive rotation curves down to 50~au scales. The
results for L1527, L1448-NB, and L1448-C are presented in
Sect.~\ref{sec:disk-candidates}. The results for the other disk
candidates are presented in Appendix~\ref{sec:other-disk-cand}.

\subsection{Disk candidates}
\label{sec:disk-candidates}

\subsubsection{L1527}
\label{sec:l1527}

L1527 is a border-line Class 0 protostar
\citep{1991ApJ...382..555L,2000prpl.conf...59A} located in the Taurus
molecular cloud at a distance of 140~pc \citep{Loinard07}\footnote{A
  recent study gives a slightly closer distance \citep[130~pc;
  ][]{2018ApJ...859...33G}. Here we adopt a distance of 140~pc for
  consistency with previous studies of L1527.}. The protostar internal
luminosity is 0.9~L$_\odot$ (Ladjelate et al., in prep.) and its
envelope mass is 1.2~M$_\odot$ \citep{2001A&A...365..440M}. The
blue-shifted lobe PA of L1527 jet is 90\degr{} (Podio et al. in
prep.). L1527 is the first Class 0 protostar in which a Keplerian disk
was claimed. \citet{Tobin12b} observed the \tcoto{} line emission with
the CARMA interferometer and suggest that the rotation profile of the
inner envelope is consistent with Keplerian rotation
($v \propto r^{-0.5}$). \citet{Yen13} observed the \ceoto{} line
emission using the SMA interferometer and find that the rotation
profile is best-fitted with $v \propto r^{-1}$ suggesting rotation and
infall with conservation of the angular momentum.  \citet{Sakai14}
used ALMA cycle 0 observations of SO and
cyclic-C$_\text{3}$H$_\text{2}$ line emission to find that SO is only
present inside a 100~au region in radius, while the
cyclic-C$_\text{3}$H$_\text{2}$ is present only outside. They
interpret these observations as a change of the gas chemical
composition at the centrifugal barrier, where the transition between
the rotating and infalling envelope and the disk is expected to occur.
\citet{Ohashi14} used ALMA cycle 0 observations of the \ceoto{} line
emission to study the envelope or disk kinematics. They find a
rotation profile of $v \propto r^{-1.2}$ from the \ceoto{} line down
to 50~au scales but suggest a possible transition to Keplerian
rotation at smaller scales. Using ALMA cycle 1 observations,
\citet{2017ApJ...849...56A} have recently confirmed Keplerian rotation
at $r < \rev{56}$~au. \rev{The disk of L1527 is nearly edge-on
  \citep{Tobin08,2015ApJ...812...59O}.}

The bottom row of Fig.~\ref{momentmaps-1} shows the 1.4~mm continuum
emission map together with the \tcoto{}, \ceoto{} and \sofs{} zeroth
and first-order moment maps in L1527. The continuum emission is
resolved spatially, and it appears to be slightly elongated along a
north-south axis. The \tcoto{}, \ceoto{}, and \sofs{} line emissions
are all well detected and are spatially coincident with the continuum
emission. The \ceoto{} emission is more extended than the \tcoto{}
emission. \rev{The \tcoto{} is} more extended than the \sofs{}
emission. The emission of each line is extended along a north-south
axis and has a well-defined velocity gradient. For all lines, the
gradients are oriented along an axis roughly orthogonal to the
red-shifted lobe of the jet, with $\Delta \theta$ of 5\degr{},
36\degr{}, and 12\degr{} for the \tcoto{}, \ceoto{} and \sofs{} lines,
respectively.

The orientation of the gradients suggests that these lines probe the
rotation of the envelope or the disk. To better quantify this
rotation, we build a rotation curve from each line. Because the size
of \sofs{} emission is only slightly larger than the size of the
synthesized beam, we construct the rotation curves from the measured
visibilities rather than the \rev{image} cubes or the PV diagrams
\citep[as done, e.g., by ][]{Yen15a}. This technique, which was also
used by \citet{Codella14} in their study of HH212 \citep[see
also][]{2014A&A...566A..74L}, allows us to constrain the velocity
curve at scales smaller than the synthesized beam and it is also
insensitive to potential deconvolution artifacts. A test of the
robustness of this technique in recovering accurate velocity profiles
is presented in Appendix~\ref{sec:benchm-rotat-curve}.  In practice,
we fit the position of the line emission in the $uv$ plane with a
circular Gaussian\footnote{We also tried to fit the visibilities in
  each channel with an elliptical Gaussian and the fitted positions
  were found to be in good agreement with those obtained with a
  circular Gaussian fit.}. Because we are interested in the disk
emission rather than the emission of the envelope or the outflow, we
ignore channels where the FWHM emission size is greater than
2\arcsec{} and channels with a centroid position that is farther away
than 1.5\arcsec{} from the continuum peak. In addition, we ignore
channels with centroid positions that are not along the assumed disk
major axis, within 3$\sigma$. For this source, we assume a disk major
axis PA of 1.5\degr{} \citep{2017ApJ...849...56A}.

\begin{table*}
  \begin{center}
    \caption{Results of the rotation curve fits for the disk candidates. \label{tab:velocityfit}}
    \begin{tabular}{llllllll}
      \hline
      \hline
      Source & PA\tablefootmark{a} & \vsys{}\tablefootmark{b} & Line & \vthau{} & $\beta$ & $\redchi$\tablefootmark{c} \\
             & (\degr{})             & (\kms{})                 &      & (\kms{}) &         &                            \\
      \hline
      L1448-NB   & 29.5 & 4.5       & \tcoto{} & 1.35 $\pm$ 0.53       & -1.25 $\pm$ 1.48       & 0.38                 \\
                 &      &           & \ceoto{} & 1.66 $\pm$ 0.11       & -1.15 $\pm$ 0.20       & 0.05                 \\
      \hline
      L1448-C    & -107 & 5.2       & \tcoto{} & 0.88 $\pm$ 0.76       & -0.44 $\pm$ 0.75       & ...\tablefootmark{d} \\
                 &      &           & \ceoto{} & 0.62 $\pm$ 0.03       & -0.67 $\pm$ 0.09       & 1.11                 \\
      \hline
      IRAS2A     & 107  & 7.5       & \ceoto{} & 1.16 $\pm$ 0.04       & -0.18 $\pm$ 0.09       & ...\tablefootmark{d} \\
      \hline
      SVS13B     & 77   & \rev{8.4} & \ceoto{} & \rev{0.42 $\pm$ 0.07} & \rev{-0.49 $\pm$ 0.28} & ...\tablefootmark{d} \\
      \hline
      IRAS4B     & -103 & 6.7       & \ceoto{} & 0.40 $\pm$ 0.12       & -2.09 $\pm$ 1.78       & ...\tablefootmark{d} \\
      \hline
      L1527      & 1.5  & 5.9       & \tcoto{} & 0.56 $\pm$ 0.04       & -1.11 $\pm$ 0.07       & 1.77                 \\
                 &      &           & \ceoto{} & 0.25 $\pm$ 0.04       & -1.88 $\pm$ 0.17       & 3.92                 \\
                 &      &           & \sofs{}  & 0.88 $\pm$ 0.10       & -0.57 $\pm$ 0.12       & 1.64                 \\
      \hline
      SerpS-MM18 & 82   & 7.9       & \ceoto{} & ...\tablefootmark{e}  & ...                    & ...                  \\
      \hline
    \end{tabular}
    \tablefoot{
      \tablefoottext{a}{Assumed disk major axis position angle for the fit.}
      \tablefoottext{b}{Assumed disk systemic velocity (in the LSR).}
      \tablefoottext{c}{Reduced $\chi^{2}$.}
      \tablefoottext{d}{$\redchi$ is undefined because the fit is based on only two points.}
      \tablefoottext{e}{No velocity curve could be fitted (see \rev{Appendix~\ref{sec:serps-mm18}}).}
    }
  \end{center}
\end{table*}

Figure~\ref{fig:uvfit-l1527} shows for each line the centroid position
in each channel, together with the zeroth-order moment map.  We see
that the centroid positions for each line are relatively well aligned
with the assumed disk major axis. The velocity gradient is also
apparent in this figure: for all lines, the centroids of the
blue-shifted channels are located south of the continuum peak
position, while the centroids of the red-shifted channels are located
north of it. To determine the rotation curves for each line, we
compute the offset from the source center by projecting the centroid
position in each channel onto the assumed disk
axis. Figure~\ref{fig:velocityfit-l1527} shows the position offset as
a function of the absolute value of the velocity difference from the
systemic velocity $v_\mathrm{sys}$. For the latter, we adopt the same
value as \citet{Ohashi14}, i.e.\rev{,} 5.9 \kms{}. In this figure, we
see that for both the \tcoto{} and \ceoto{} lines, the position offset
from the continuum peak decreases while $|v - v_\mathrm{sys}|$
increases, indicating that the gas probed by these lines rotates
faster as it gets closer to the continuum peak. The \sofs{} line shows
the same trend for high-velocity channels.  For low-velocity channels,
the position offset increases roughly linearly as
$|v - v_\mathrm{sys}|$ increases, as also observed by
\citet{Ohashi14}.

Then, we fit the velocities as a function of position with a
power-law:

\begin{equation}
  v \left( r \right) - v_\mathrm{sys} = \vthau \,
  {\left( \frac{p \,d} {\mathrm{200 \, au}} \right)}^{\beta},
\end{equation}

\noindent
where $p$ is the position offset in arcseconds (counted as positive on
the red-shifted side of the emission, and as negative on the other
side), $d$ is the source distance in pc, $\beta$ is the power-law
exponent, and \vthau{} is the projected velocity at 200~au from the
center. Both $\beta$ and \vthau{} are left as free parameters. For the
\sofs{} fit, we ignore channels between 4.8 and 7.2~\kms{} because, as
mentioned above, \rev{the} low-velocity channels cannot be fitted with
the same velocity curve as \rev{the} high-velocity channels.

The best-fit results are shown as solid lines in
Fig.~\ref{fig:velocityfit-l1527}. The best-fit values of $\vthau$ and
$\beta$, and the reduced $\chi^2$, are given in
Table~\ref{tab:velocityfit}. The best-fit value of $\beta$ is
$-1.11 \pm 0.07$, $-1.88 \pm 0.17$, and $-0.57 \pm 0.12$ for the
\tcoto{}, \ceoto{}, and \sofs{} lines, respectively. Therefore, the
velocity curve we derive from the \sofs{} line emission is consistent
with Keplerian rotation ($\beta = -0.5$). Fitting the \sofs{} position
in each channel with a Keplerian law:

\begin{equation}
  v \left( r \right) - v_\mathrm{sys} = \sqrt{G \, \Mstar{}} \,
  \sin{i} \, {\left( p \, d \right)}^{-0.5},
\end{equation}

\noindent
where $G$ is the gravitational constant, $\Mstar{}$ is the central
mass and $i$ is the disk inclination, we obtain
$\Msinisqr{} = 0.20 \pm 0.01$~\Msun{}. 

In Fig.~\ref{fig:velocityfit-l1527}, we also show the velocity curve
obtained by \citet{2017ApJ...849...56A} from \ceoto{}
observations. These authors derive a power-law exponent of -0.50 at
radii smaller than 56~au, and -1.22 beyond. The double power-law
agrees well with our \tcoto{} and \ceoto{} observations. However, our
\sofs{} observations are better reproduced with a single power law or
a Keplerian law for a central mass of $\Msinisqr{} =
0.20$~\Msun{}. The lowest velocity channel that is consistent with the
Keplerian velocity curve is at a position offset of about
$0.65\arcsec$, that is\rev{,} a radius of 90~au.

In Fig.~\ref{fig:pv-l1527}, we show position-velocity (PV) cuts
through the position of the L1527 continuum peak along the assumed
disk axis, for the \tcoto{}, \ceoto{}, and \sofs{} lines. The PV
diagrams for the \tcoto{} and \ceoto{} lines indicate both infall and
rotation \citep[see\rev{,} e.g.\rev{,}][]{2012ApJ...748...16T}. On the
other hand, the PV diagram for the \sofs{} is consistent with rotation
alone.  On these diagrams, we also show the power-law velocity curves
we obtain for each of these lines with a fit in the $uv$ plane. We see
that the velocity curves provide a good fit to the emission peak in
each velocity channel.  For each diagram, we also plot \rev{the
  outermost Keplerian curve that is tangential to the first emission
  contour, with a 6$\sigma$ threshold\footnote{\rev{For a 3$\sigma$
      threshold, we obtain slightly larger masses ($\sim30\%$):
      $\Msinisqr = 0.9$~\Msun{}, 0.8~\Msun{} and 0.5~\Msun{} for the
      \tcoto{}, \ceoto{} and \sofs{} lines, respectively.}}
  \citep[see][for a justification of the
  method]{2016MNRAS.459.1892S,2018ApJ...860..119G}}. We obtain
$\Msinisqr = 0.7$~\Msun{}, 0.6~\Msun{} and 0.4~\Msun{} for the
\tcoto{}, \ceoto{}, and \sofs{} lines, respectively.

To summarize our findings for this source, we find evidence for
Keplerian rotation, but with the \sofs{} line only. Both the \tcoto{}
and \ceoto{} line emission appear to be dominated by the rotation and
infall of the envelope. Keplerian rotation is observed with the
\sofs{} line at $r \le 90$~au. As discussed in
Appendix~\ref{sec:benchm-rotat-curve}, a Keplerian fit in the $uv$
plane gives a lower limit on the mass $\Mstar{}$ of the central
object, while a fit of the first contours in PV diagrams gives an
upper limit. By combining these two techniques and assuming a disk
inclination of $85\degr$ \citep{Tobin08}, we obtain
$\Msinisqr \simeq \Mstar{} = 0.2 - 0.4 \, \Msun{}$.

\subsubsection{L1448-NB}

\begin{figure*}
  \sidecaption
  \includegraphics[width=12cm]{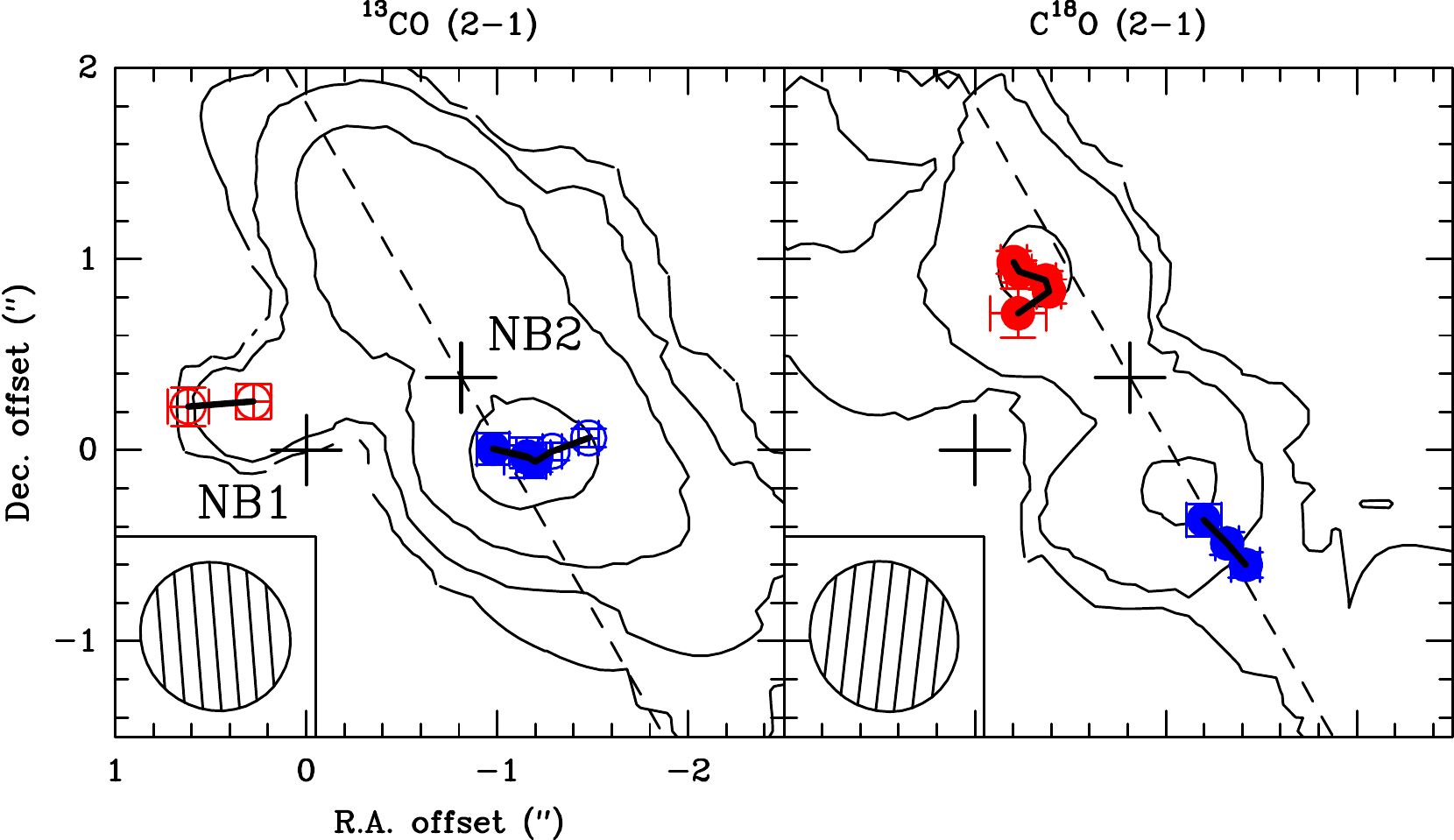}
  \caption{Same as in Fig.~\ref{fig:uvfit-l1527} for L1448-NB. Black
    crosses show the positions of the NB1 and NB2 continuum
    peaks. \label{fig:uvfit-l1448nb}}
\end{figure*}

\begin{figure*}
  \sidecaption
  \includegraphics[width=12cm]{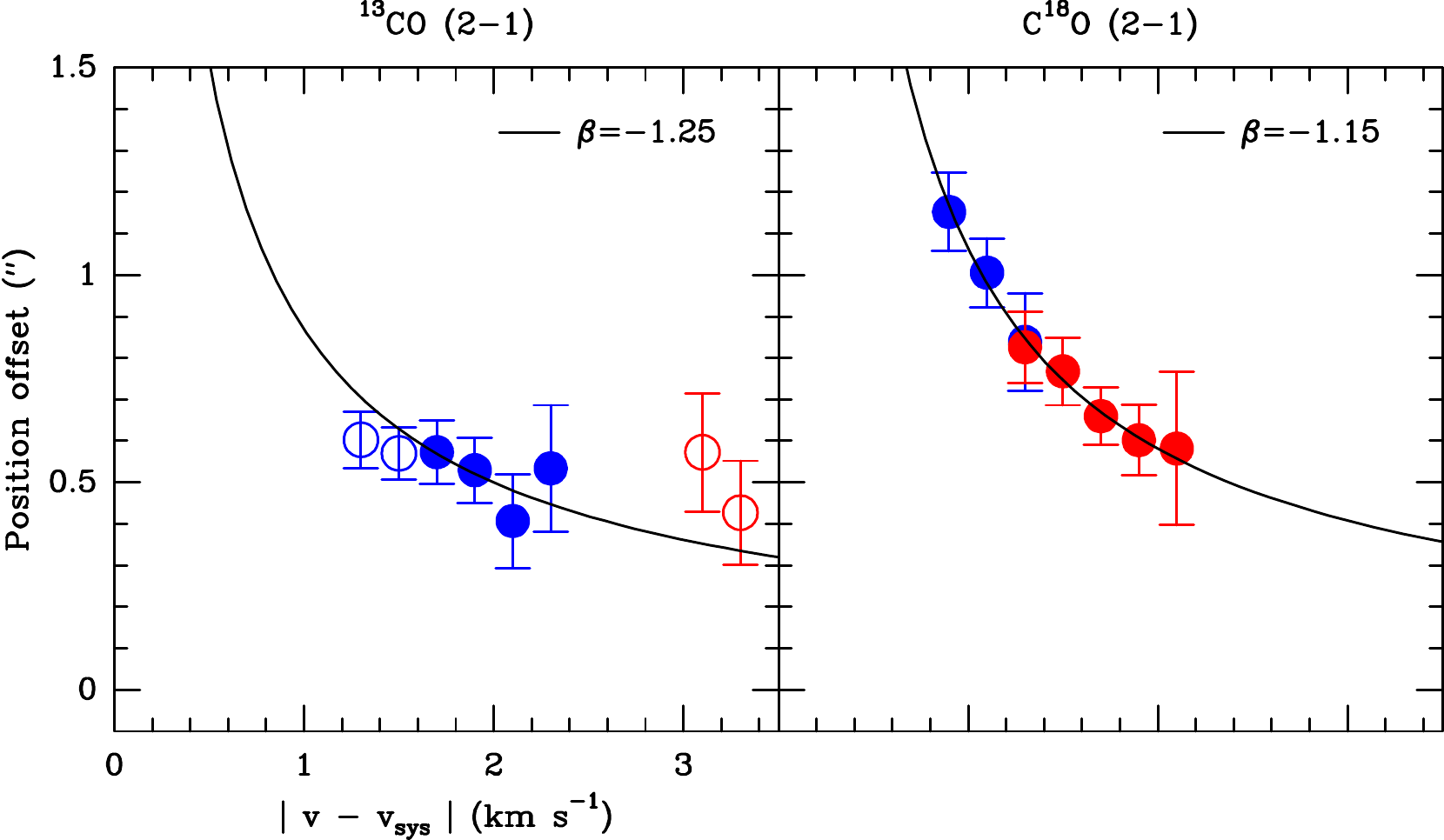}
  \caption{Same as in Fig.~\ref{fig:velocityfit-l1527} for
    L1448-NB. The reference position is
    NB2. \label{fig:velocityfit-l1448nb}}
\end{figure*}

\begin{figure*}
  \sidecaption
  \includegraphics[width=12cm]{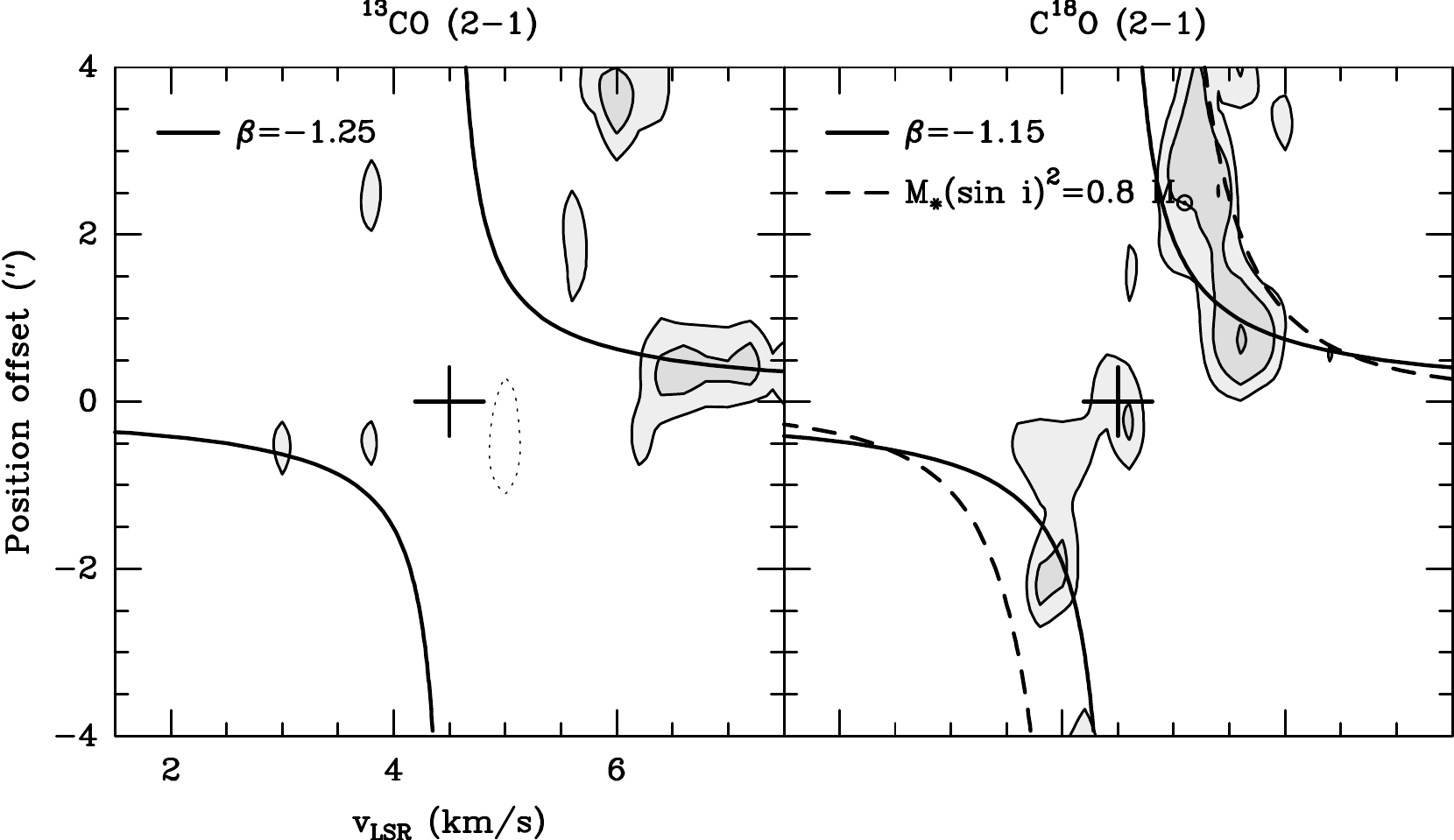}
  \caption{Same as in Fig.~\ref{fig:pv-l1527} for L1448-NB. Dashed
    curve shows a Keplerian curve for $\Msinisqr = 0.8~\Msun$. The
    reference position is NB2. \label{fig:pv-l1448nb}}
\end{figure*}

L1448-NB (also known as L1448-IRS3B and Per-emb-33) is a Class 0
protostar \citep{1990ApJ...365L..85C} located in the L1448-N complex
in Perseus, at a distance of ~293~pc \citep{2018ApJ...865...73O}. This
source is a triple system, composed of L1448-NB1 (Per-emb-33 A) and a
close binary, L1448-NB2. These two are separated by 260~au
\citep[0.9\arcsec;][]{Tobin15,2016Natur.538..483T,2019A&A...621A..76M}. The
two components of the binary (Per-emb-33 B and Per-emb-33 C) are
separated by 80~au \citep[0.26\arcsec; ][]{Tobin16a} and they are
unresolved with our observations. The internal luminosity of the
triple system is 3.9~L$_\odot{}$ (Ladejate et al. in prep.) and it is
surrounded by a 4.8~M$_\sun$ envelope \citep{Sadavoy14}. The jet
position angle is -80\degr{} (Podio et al. in prep). \cite{Yen15a}
observed a large velocity gradient in this source orthogonal to the
jet. They attribute this gradient to rotational and infall motions and
derive a centrifugal radius between 410 and 900 au, that is\rev{,} 1.4
and 3\arcsec. Recent ALMA 1.3 mm continuum observations have suggested
the presence of one or two spiral arms that appear to originate from
NB2 and extend to NB1 \citep{2016Natur.538..483T}. The same authors
observed a rotation pattern in the \ceoto{} line emission that they
interpret as Keplerian rotation of a disk around NB2, implying a
central stellar mass\footnote{\cite{2016Natur.538..483T} assumed a
  distance of 235~pc \citep{Hirota08}. For the distance adopted here
  (293~pc), the central stellar mass is $\sim 1.2$~\Msun{}} of
$\sim 1$~\Msun{}. In their scenario, NB1 would be the result of the
gravitational fragmentation of the disk. However,
\citet{2019A&A...621A..76M} find no evidence of continuum emission
tracing a disk centered on NB2, and they suggest that the spiral
structures observed by \citeauthor{2016Natur.538..483T} trace tidal
arms due to the gravitational interaction between the two components
of the binary.

First-order moment maps of the \tcoto{}, \ceoto{}, and \sofs{}
emission in L1448-NB are shown in Fig.~\ref{momentmaps-1}. Both the
\tcoto{} and \ceoto{} line emissions are centered around NB2 and are
elongated approximately along an axis orthogonal to the jet direction,
with a well-defined velocity gradient. No \sofs{} emission is detected
in this source at a 3$\sigma$ level. Fitting the first-order moment
maps, we measure a gradient PA of $(59 \pm 7)$\degr{} and
$(44 \pm 5)$\degr{} for the \tcoto{}, and \ceoto{} lines
respectively. These correspond to angles with respect to the
\rev{direction perpendicular to the} jet of 49\degr{} and 34\degr{},
respectively.

Figure~\ref{fig:uvfit-l1448nb} shows the centroid positions in each
velocity channel, together with the zeroth-order moment maps for the
\tcoto{} and \ceoto{} lines. In this figure, we also show the disk
orientation measured by
\citet[PA~=~29.5\degr]{2016Natur.538..483T}. The disk is assumed to
be centered on NB2 because the line emission is centered on this
source. The centroids of the \tcoto{} line are not aligned with the
assumed disk axis, probably because of contamination by the
outflow. On the contrary, the \ceoto{} centroids are located close to
the disk axis. The corresponding rotation curves are shown in
Fig.~\ref{fig:velocityfit-l1448nb}. To build the rotation curves, we
adopt a systemic velocity\footnote{The choice of the systemic velocity
  is justified a posteriori by the red and blue points in the right
  panel of Fig.~\ref{fig:velocityfit-l1448nb}, which are consistent
  with a single rotation curve. If the assumed systemic velocity were
  incorrect, the blue and red points would not overlap, and they could
  not be fitted with a rotation curve.} of 4.5~\kms{}. The best-fit
parameters for the rotation curve are given in
Table~\ref{tab:velocityfit}. We find that the \tcoto{} emission is
best fitted with $\beta = -1.25 \pm 1.48$. This fit is uncertain
because it is based on four channels only; other channels (marked with
open symbols in Fig.~\ref{fig:uvfit-l1448nb} and
\ref{fig:velocityfit-l1448nb}) are not considered in the fit because
their centroids are not aligned with the assumed disk axis. Using the
\ceoto{} line emission, we find a best-fit $\beta = -1.15 \pm 0.20$
for position offsets greater than $\sim 0.6\arcsec$, that is\rev{,}
$r > 175$~au.

PV diagrams along the assumed disk axis are shown in
Fig.~\ref{fig:pv-l1448nb}. The PV diagram for the \tcoto{} emission
has a low signal-to-noise ratio and, therefore, it is difficult to
interpret. The PV diagram for the \ceoto{} emission has a higher
signal-to-noise ratio and it is consistent with rotation alone
\rev{(i.e., with no infall)}. Fitting the first contour of the
\ceoto{} emission with a Keplerian curve, we find
$\Msinisqr = 0.8~\Msun$, i.e.\rev{,} $\Mstar = 1.4~\Msun$, assuming
$i = 45.4\degr{}$ \citep{2016Natur.538..483T}. We note that the curve
fits the first contour of red-shifted \ceoto{} emission quite well,
but not that of the blue-shifted emission.

To summarize, we find no evidence of Keplerian rotation in this
source. Instead, the rotation profile derived from the \ceoto{} line
is consistent with $\beta = 1$ for $r > 175$~au. We conclude that if a
Keplerian disk is present in this source, its radius is smaller than
175~au.

\subsubsection{L1448-C}

L1448-C (also known as L1448-mm or Per-emb-26) is another Class 0
protostar \citep{1989ApJ...341..208A} located in the L1448-N complex
in Perseus. This source is a multiple system, but with a large
separation: its companion, L1448-C S (Per-emb-42), is located
8\arcsec{} to the southeast
\citep[]{Jorgensen06b,Tobin07,2019A&A...621A..76M}. L1448-C itself
appears as a single source at scales \rev{down to} 50~au
\citep{2019A&A...621A..76M}. Its internal luminosity is 10.9~L$_\odot$
and its envelope mass is 2.0~M$_\odot$ \citep[Ladejate et al. in prep;
][]{Sadavoy14}. L1448-C drives a prominent outflow with multiple
high-velocity knots seen in CO, SO and H$_{2}$O lines
\citep{Hirano10,Kristensen11}. The blue-shifted lobe of the jet has a
position angle of -17\degr{} (Podio et al. in prep) and an inclination
with respect to the plane of the sky of 21\degr{}
\citep{Girart01}. From CARMA 1.3~mm observations, \citet{Tobin15} find
that the L1448-C continuum is elongated along a direction
perpendicular to the jet, perhaps indicative of a disk-like
structure. \citet{Yen13,Yen15a} measure a velocity gradient with a PA
of 200\degr{}, dominated by the outflow component. Fitting only the
component perpendicular to the jet, they find that the rotation
profile was consistent with rotation with conservation of the angular
momentum and they derive a centrifugal radius of 170-200~au.

In our maps, both the \tcoto{} and \ceoto{} line emissions are
extended along an axis close to the normal direction of the jet (see
Fig.~\ref{momentmaps-1}). No \sofs{} emission is detected in the
observed velocity range. Clear velocity gradients are observed in
\tcoto{} and \ceoto{} line mean velocity maps, and the fitted PA are
$(-150 \pm 15)$~\degr{} and $(-111 \pm 6)$~\degr{} for the \tcoto{}
and \ceoto{} lines, respectively. The corresponding values of
$\Delta \theta$ are 43 and 4\degr{} for the \tcoto{} and \ceoto{}
lines, respectively.

\begin{figure*}
  \sidecaption
  \includegraphics[width=12cm]{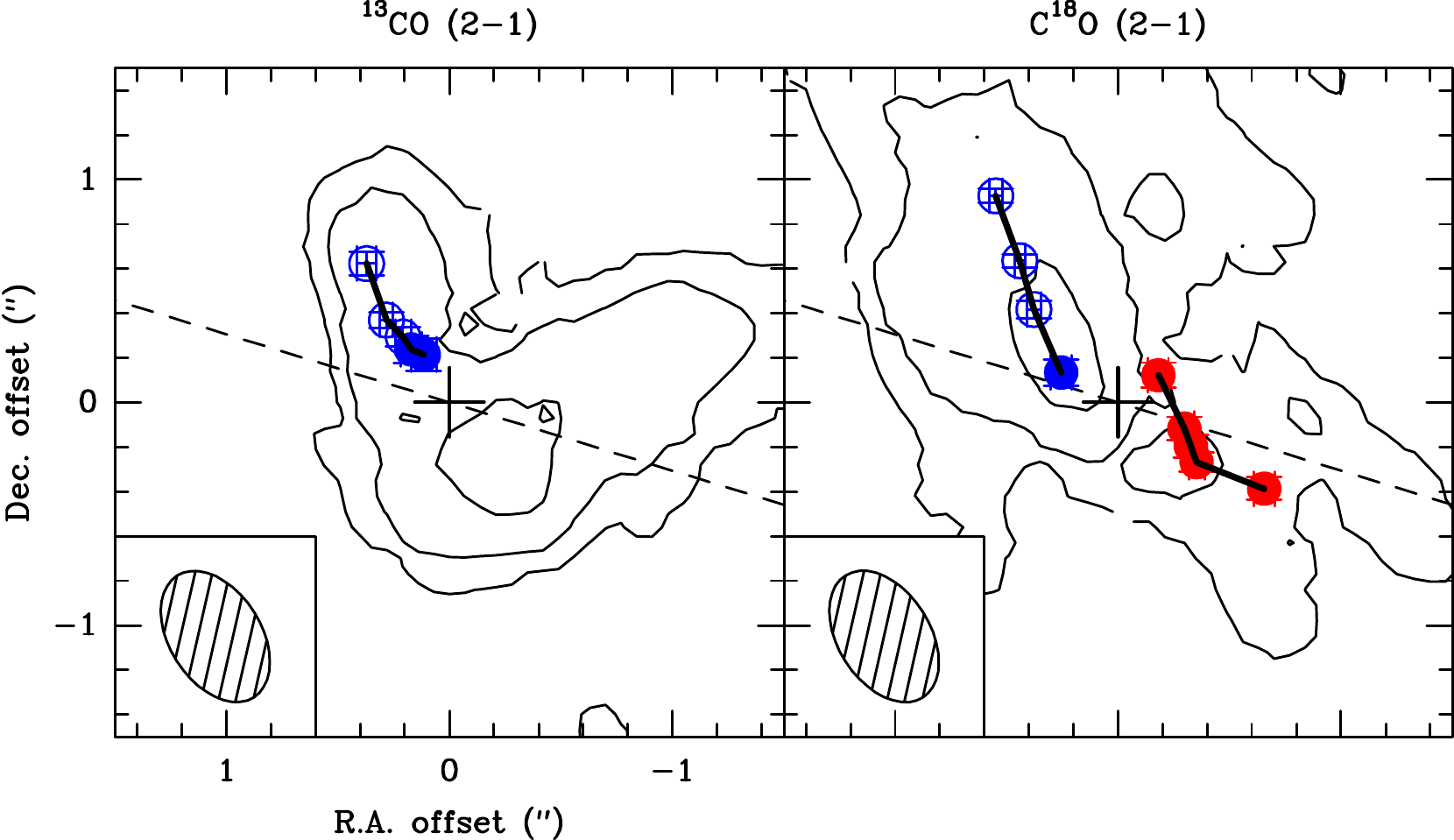}
  \caption{Same as in Fig.~\ref{fig:uvfit-l1527} for
    L1448-C.\label{fig:uvfit-l1448c}}
\end{figure*}

\begin{figure*}
  \sidecaption
  \includegraphics[width=12cm]{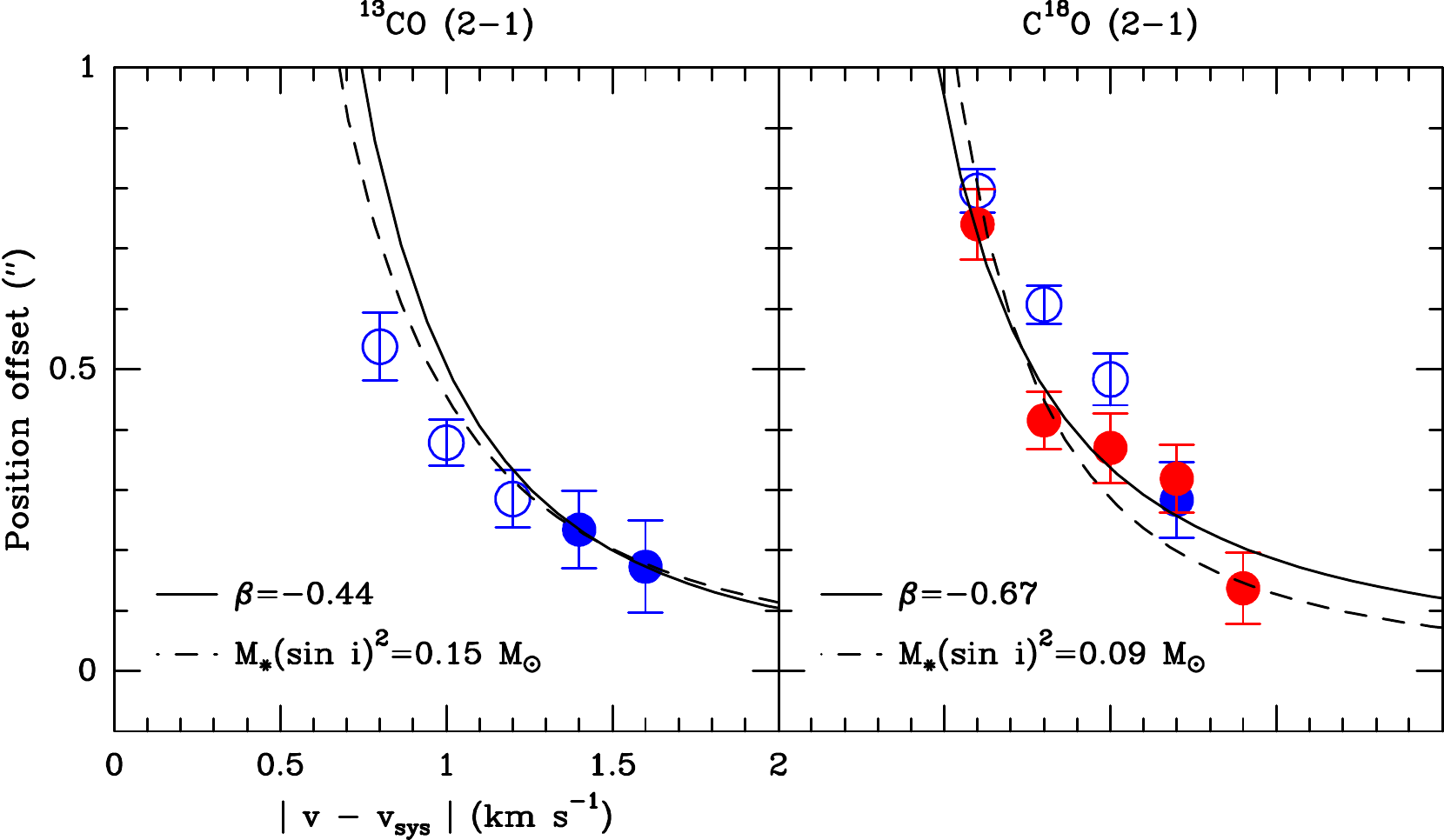}
  \caption{Same as in Fig.~\ref{fig:velocityfit-l1527} for
    L1448-C. The dashed lines show the result of a Keplerian fit with
    $\Msinisqr = 0.15$~\Msun{} and 0.09~\Msun{} for the \tcoto{} and
    \ceoto{} lines, respectively.\label{fig:velocityfit-l1448c}}
\end{figure*}

\begin{figure*}
  \sidecaption
  \includegraphics[width=12cm]{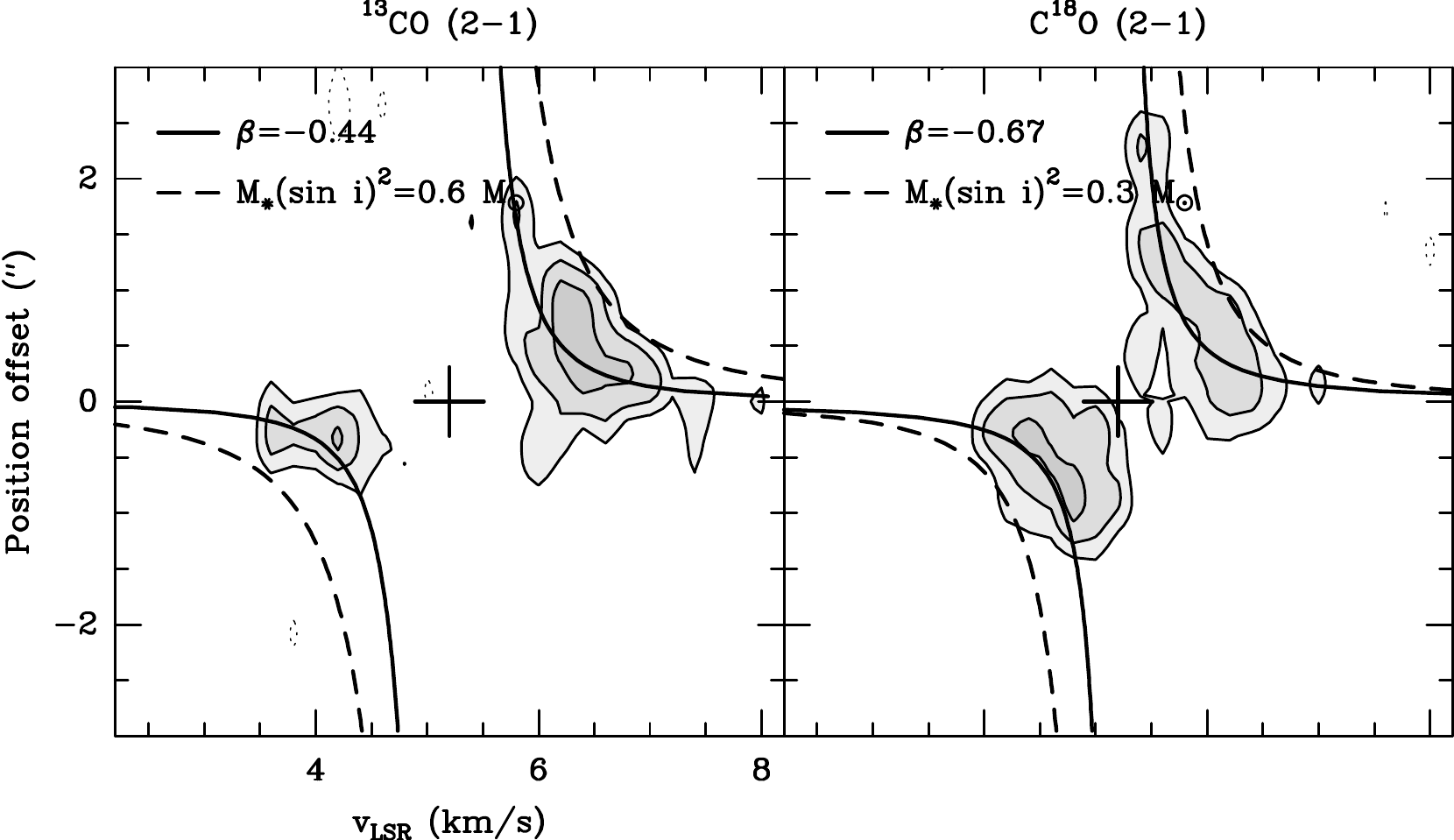}
  \caption{Same as in Fig.~\ref{fig:pv-l1527} for L1448-C. The dashed
    curves shows Keplerian velocity profiles with $\Msinisqr = 0.6~\Msun$
    and $\Msinisqr = 0.3~\Msun$ for the \tcoto{} and \ceoto{} lines,
    respectively. \label{fig:pv-l1448c}}
\end{figure*}

The results of the $uv$ fit for these two lines are shown in
Fig.~\ref{fig:uvfit-l1448c}, and the corresponding rotation curves are
shown in Fig.~\ref{fig:velocityfit-l1448c}. Here we \rev{use} a
systemic velocity\footnote{\rev{The systemic velocity is estimated
    from the \tcoto{} and \ceoto{} PV diagrams (see
    Fig.~\ref{fig:pv-l1448c}) which are centered around 5.2~\kms{}.}}
of 5.2~\kms{} and a disk position angle of -107\degr{}, meaning that
it is orthogonal to the jet axis.  For the \tcoto{} emission, we
measure the centroid positions only in a few blue-shifted channels;
red-shifted channels are contaminated by the outflow emission. Only
two channels have centroids that are located along the disk axis. For
the \ceoto{} emission, we measure the centroid positions of 9
channels, but several blue-shifted channels have centroids that are
not aligned with the disk axis. For the \tcoto{} emission, we find
that the best-fit $\beta$ value is $-0.44 \pm 0.75$, while for the
\ceoto{}, we find $\beta = -0.67 \pm 0.09$. Both rotation curves are
consistent with Keplerian rotation, but the uncertainty on the $\beta$
index for the \tcoto{} line is quite large because the fit is based on
two channels only. The rotation curve we derive with the \ceoto{} line
is much better constrained, and the best-fit value of $\beta$ is close
to the value expected for Keplerian rotation (-0.5). From a fit with a
Keplerian law, we obtain $\Msinisqr = 0.15 \pm 0.03$~\Msun{} and
$\Msinisqr = 0.09 \pm 0.01$~\Msun{} for the \tcoto{} and \ceoto{}
rotation curves, respectively.

Figure~\ref{fig:pv-l1448c} shows the PV diagrams along the assumed
disk axis. Both diagrams are consistent with rotation alone. Fitting
the first emission contour with a Keplerian law, we obtain
$\Msinisqr = 0.6~\Msun{}$ and $\Msinisqr = 0.3~\Msun{}$ for the
\tcoto{} and \ceoto{} lines, respectively.

To summarize our finding for this source, we derive a velocity curve
with the \ceoto{} line that is close to Keplerian rotation. The
velocity profile we derive with the \tcoto{} line is also consistent
with Keplerian rotation, but it has a large uncertainly. Keplerian
rotation is observed with the \ceoto{} line at $r < 0.7\arcsec$, that
is, $< 200$~au. Combining the central mass obtained from the fit in
the $uv$ plane and from the PV diagram of the \ceoto{} line, we find
$\Msinisqr \simeq \Mstar{} = 0.1 - 0.3 \, \Msun{}$. Assuming that the
disk has the same inclination as the jet
\citep[$i = 21\degr{}$][]{Girart01}, this gives
$\Mstar = 0.8-2.3$~\Msun{}.

\section{Discussion}
\label{sec:discussion} 

\begin{table*}
  \begin{center}
    \caption{Overview of evidence of the presence of disks in the
      CALYPSO sample of Class 0 objects. \label{tab:disk-evidence-summary}}
    \begin{tabular}{lllllllllllll}
      \hline
      \hline
      Source & & \multicolumn{2}{c}{Cont. emission\tablefootmark{a}} & & \multicolumn{2}{c}{Ang. momentum\tablefootmark{b}} & & \multicolumn{4}{c}{Kinematics on small scales\tablefootmark{c}} \\
      \cline{3-4}
      \cline{6-7}
      \cline{9-12}
      &  & Compact                 & Cont.                 &  & Const.               & $<j_{100~\mathrm{au}}>$\tablefootmark{g} &  & Grad.\tablefootmark{h} & Kepl.                 & Kepl.                   & \Mstar{}\tablefootmark{k} \\
      &  & cont.                   & size\tablefootmark{e} &  & $j$\tablefootmark{f} & ($10^{-4}$                               &  &                        & rot.\tablefootmark{i} & radius\tablefootmark{j} & (\Msun{})                 \\
      &  & source\tablefootmark{d} & (au)                  &  &                      & km~s$^{-1}$~pc)                          &  &                        &                       & (au)                    &                           \\
      \hline
      L1448-2A       &  & \xmark & $< 50$                   &  & \cmark & $4.5 \pm 0.2$  &  & \xmark & ...                     & ...     & ...               \\
      L1448-NB       &  & \xmark & $< 300$\tablefootmark{l} &  & \cmark & $16.0 \pm 0.4$ &  & \cmark & \xmark                  & $< 175$ & \rev{...}         \\
      L1448-C        &  & \cmark & $46 \pm 15$              &  & \cmark & $6.0 \pm 0.2$  &  & \cmark & \cmark                  & 200     & $0.8-2.3$         \\
      NGC1333-IRAS2A &  & \cmark & $< 62$                   &  & \cmark & $3.8 \pm 0.4$  &  & \cmark & \xmark                  & ...     & ...               \\
      SVS13B         &  & \cmark & $< 75$                   &  & \cmark & $2.5 \pm 0.2$  &  & \cmark & \cmark\tablefootmark{m} & 150     & \rev{$0.05-0.12$} \\
      NGC1333-IRAS4A &  & \xmark & $< 94$                   &  & \xmark & ...            &  & \xmark & ...                     & ...     & ...               \\
      NGC1333-IRAS4B &  & \cmark & $156 \pm 31$             &  & \cmark & $2.5 \pm 0.3$  &  & \cmark & \xmark                  & ...     & ...               \\
      IRAM04191      &  & \cmark & $< 60$                   &  & \xmark & ...            &  & \xmark & ...                     & ...     & ...               \\
      L1521F         &  & \cmark & $< 60$                   &  & \xmark & ...            &  & \xmark & ...                     & ...     & ...               \\
      L1527          &  & \cmark & $54 \pm 10$              &  & \cmark & $5.6 \pm 0.1$  &  & \cmark & \cmark                  & 90      & $0.2 -0.4$        \\
      SerpM-S68N     &  & \xmark & $< 53$                   &  & \xmark & ...            &  & \xmark & ...                     & ...     & ...               \\
      SerpM-SMM4     &  & \cmark & $305 \pm 42$             &  & \xmark & ...            &  & \xmark & ...                     & ...     & ...               \\
      SerpS-MM18     &  & \cmark & $< 61$                   &  & \xmark & ...            &  & \cmark & \xmark                  & ...     & ...               \\
      SerpS-MM22     &  & \cmark & $88 \pm 14$              &  & \xmark & ...            &  & \xmark & ...                     & ...     & ...               \\
      L1157          &  & \cmark & $< 70$                   &  & \xmark & ...            &  & \xmark & ...                     & ...     & ...               \\
      GF9-2          &  & \cmark & $26 \pm 9$               &  & \cmark & $3.9 \pm 0.3$  &  & \xmark & ...                     & ...     & ...               \\
      \hline
    \end{tabular}
    \tablefoot{
      \tablefoottext{a}{Compact continuum emission
        \citep{2019A&A...621A..76M}.}
      \tablefoottext{b}{Specific angular momentum profile
        (\citeauthor{Gaudel2020} subm.).}
      \tablefoottext{c}{Gas kinematics down to 50~au scales (this
        study).}
      \tablefoottext{d}{Detection of a compact continuum component
        \rev{in addition to emission of the extended envelope}. The
        \xmark{} and \cmark{} symbols indicate a non-detection and a
        detection, respectively.}
      \tablefoottext{e}{Radius of the compact continuum component
        scaled to the assumed distance.}  \tablefoottext{f}{Constant
        specific angular momentum $j$ at scales between 50 and
        1200~au.}
      \tablefoottext{g}{Mean value of the specific angular momentum at
        scales $< 100$~au, for sources in which $j$ is constant.}
      \tablefoottext{h}{Velocity gradient perpendicular to the jet,
        \rev{within $\pm 45\degr{}$}.}
      \tablefoottext{i}{Detection of Keplerian rotation.}
      \tablefoottext{j}{Radius of the region in Keplerian rotation.}
      \tablefoottext{k}{Mass of the central object.}
      \tablefoottext{l}{Upper limit derived from a model centered on
        NB2. There is no clear evidence for a disk-like component
        centered on NB2 at ${\buildrel < \over \sim} 1\arcsec{}$
        scales \cite[][Table C3]{2019A&A...621A..76M}.}
      \tablefoottext{m}{Tentative detection.}
    }
  \end{center}
\end{table*}

\subsection{Evidence for disks in the CALYPSO sample}

Our observations and analysis show that about half of the protostars
of our sample (7 out of 16) have velocity gradients oriented close to
the normal direction of the jet axis, at a few hundred au
scales. These gradients are interpreted as rotation of the envelope or
the disk about the jet axis. Among these sources, we detect Keplerian
rotation in two sources, L1527 and L1448-C. In a third source, SVS13B,
the rotation curve we derive is also consistent with Keplerian
rotation, but as noted in Appendix~\ref{sec:svs13b}, the rotation
curve for this source is very uncertain and we do not discuss it
further. In L1527 and L1448-C, we estimate Keplerian radii of 90 and
200~au, and central object masses of $0.2-0.4$ and $0.8-2.3~\Msun{}$,
respectively.

Our results are summarized in
Table~\ref{tab:disk-evidence-summary}. In this Table we also report
the results from \citet{2019A&A...621A..76M}, who have recently used
CALYPSO observations of the continuum emission at 93~GHz and 230~GHz
to determine if disk-like structures are present in Class 0 protostars
of the sample. For this, they have modeled the visibilities with a
Plummer-like envelope density profile and an additional Gaussian
component to mimic the disk emission. They find that a Gaussian
component is needed in 12 protostars of the sample, but these
disk-like components are spatially resolved in only six of these. The
FWHM size of resolved disk-like components varies between ~50 and
~300~au. We also report the results from
\citeauthor{Gaudel2020} (subm.) who have studied the specific angular
momentum profile in CALYPSO protostellar envelopes using
N$_2$H$^+$~(1-0) and \ceoto{} line observations. They find that the
specific angular momentum at scales between 50 and 1200~au is
conserved in eight protostars. In these protostars, the mean value of
the specific angular momentum at scales $< 100$~au (noted as
$<j_{100~\mathrm{au}}>$) is $(2-16) \times 10^{-4}$~km~s$^{-1}$~pc.

As seen in Table~\ref{tab:disk-evidence-summary}, \rev{all sources
  with a velocity gradient orthogonal to the jet also exhibit
  disk-like continuum components, except L1448-NB.} However, this
source is a binary and it is embedded in a 200~au circumbinary
structure \citep{2019A&A...621A..76M}. All the sources in which we
detect a velocity gradient orthogonal to the jet also have a constant
specific angular momentum between 50 and 1200~au, with the exception
of SerpS-MM18. Interestingly, L1527 and L1448-C, where we detect
Keplerian rotation, are also among the sources with the highest
$<j_{100~\mathrm{au}}>$ values. The only source with a higher
$<j_{100~\mathrm{au}}>$ value than L1527 and L1448-C is L1448-NB, but
as has already been mentioned, \rev{this source} is peculiar. The
Keplerian radii we derive here for L1527 and L1448-C are,
respectively, $\sim$2 and $\sim$4 times larger than the compact
continuum sources measured by \citet{2019A&A...621A..76M}. As
discussed by \citet{2019A&A...621A..76M}, their analysis of the
continuum emission is sensitive to deviation from the protostar
envelope density profile, which is presumably due to the disk. Our
results suggest that the gaseous disk may extend further away than the
compact continuum emission. Indeed, protoplanetary disks around Class
II protostars are also larger in gas emission than in millimeter
continuum emission \citep[by a factor 2 on average;
][]{2018ApJ...859...21A}.

\subsection{Rotation curves of L1527, L1448-C and L1448-NB}

In L1527, we detect Keplerian rotation with the \sofs{} line
only. Both the \tcoto{} and \ceoto{} lines indicate steeper velocity
profiles, with $\beta = -1.1 \pm 0.1$ and $\beta = -1.9 \pm 0.2$,
respectively. The velocity profiles we obtain with these two lines are
in reasonable agreement with \cite{2017ApJ...849...56A}, who derived a
$\beta = -1.22$ at radii larger than 56~au (0.4\arcsec{}) and
$\beta = -0.5$ at a smaller radius. Although the sensitivity and the
spatial resolution of our observations are not sufficient to detect
the ``kink'' in the velocity profile at 56~au, our observations are
consistent with it, as seen in
Fig.~\ref{fig:velocityfit-l1527}. \cite{2017ApJ...849...56A}
interpreted this kink as a transition region between the infalling
envelope and the Keplerian disk at $r = 74$~au\footnote{The difference
  between the apparent position of the kink and the Keplerian disk
  radius is due to the limited angular resolution and the close to
  edge-on geometry, which makes the kink to appear closer to the
  central object than the actual Keplerian disk radius
  \citep{Aso15}}. Our observations and analysis of the \sofs{}
emission show that Keplerian rotation extends up to 0.7\arcsec{} from
the continuum peak, that is\rev{,} $r \sim 100$~au (see
Fig.~\ref{fig:velocityfit-l1527}). \citet{Sakai14} studied the
kinematics of L1527 using C$_3$H$_2$ and \sofs{} line
observations. From this analysis, they estimate that the radius of the
centrifugal barrier\footnote{The radius of the centrifugal barrier is
  half the centrifugal radius \citep{Sakai14}.} is 100~au. The
Keplerian radius we derive from our \sofs{} observations and analysis
is in agreement with this value. \citet{Sakai14} and \citet{Ohashi14}
argue that the \sofs{} emission originates in a ring at the position
of the centrifugal barrier. However, observations at higher angular
resolution show that SO is also present inside the centrifugal barrier
\citep{2017MNRAS.467L..76S}.

\rev{A possible explanation for the differences between the disk
  radius derived by \cite{2017ApJ...849...56A} from \ceoto{} line
  observations and the radius we derive from \sofs{} line observations
  is the different spatial resolution of the observations. The
  synthesized beam size of the \ceoto{} line observations of
  \citeauthor{2017ApJ...849...56A} is
  $0.50\arcsec \times 0.40\arcsec$, while the synthesized beam size of
  our \sofs{} observations is $0.72\arcsec \times 0.64\arcsec$. In
  principle, if the \sofs{} line was not sufficiently resolved
  spatially, the transition between the two power-law regimes of the
  velocity profile could be blurred. However, we show in
  Appendix~\ref{sec:benchm-rotat-curve} that we can recover the
  velocity profile of a disk with a radius of 100~au at a distance of
  293~pc, which corresponds to an apparent radius of 0.34\arcsec. This
  is smaller than the disk radius derived by
  \citeauthor{2017ApJ...849...56A} (74 au at 140 pc, i.e.\rev{,}
  0.53\arcsec). Another possible explanation is the opacity of the
  \ceoto{} line.} If this line is optically thick close to the
systemic velocity, the Keplerian disk, which rotates faster than the
envelope, would be seen only in the line wings. Closer to the source
systemic velocity, the disk would be masked by the optically thick
envelope. In this scenario, the change in the velocity profile at
$r = 56$~au would be due to \rev{a change in the \ceoto{} line
  opacity}: optically thick for $| v - v_\mathrm{sys} | < 2$~\kms{},
and optically thin for $| v - v_\mathrm{sys} | \ge 2$~\kms{}. The
\sofs{} line is less affected because its critical density is higher
than the critical density of the \ceoto{} line and it is therefore
excited only in the densest part of the envelope and the
disk. \rev{Indeed, the opacity of the \ceoto{} line was estimated by
  \citet{2018A&A...615A..83V} from the \tcoto{} to \ceoto{} line
  intensity ratio. They find that the \ceoto{} line is optically thick
  in the mid-plane (see their Fig.~2).  Confirming that the change in
  power-law index of the velocity profile at 56~au is due to the
  opacity of the \ceoto{} line} would require detailed modeling of the
line radiative transfer, which is beyond the scope of the present
paper.

In L1448-C, we detect Keplerian rotation with the \ceoto{} line. This
is the first detection of the Keplerian disk in this source. Although
the rotation curve we derive with the \tcoto{} line is consistent with
Keplerian rotation, the value of the velocity profile index $\beta$
derived with this line has a large uncertainty. We estimate a
Keplerian radius of 200~au and a central stellar mass between 0.8 and
2.3~$\Msun{}$. L1448-C's disk is twice larger than L1527's disk, and
the central stellar mass is also significantly higher. Unfortunately,
the disk is only marginally resolved by our
observations. Higher-angular resolution are needed to better
characterize this disk and, in turn, to compare its properties with
that of other disks around Class 0 protostars.

Although we do not detect Keplerian rotation in L1448-NB, we briefly
discuss the rotation profile in this source, in which the presence of
a unstable disk with a radius of $\sim$400~au has recently been
claimed by \cite{2016Natur.538..483T}. In this source, our \ceoto{}
observations and modeling suggest a rotation profile with
$\beta = -1.15 \pm 0.15$ down to 0.6\arcsec{} (175~au) from the
continuum peak. The rotation profile we derive with \tcoto{} is
steeper, but the line is affected by outflow emission (see
Fig.~\ref{fig:uvfit-l1448nb}). Our analysis suggests that the disk
observed by \cite{2016Natur.538..483T} is not in Keplerian rotation at
$r > 175$~au, although it is possible that Keplerian rotation is
present on smaller scales.

\subsection{Could we have missed some of the disks?}

Our observations and analysis of the gas kinematics in the CALYPSO
sample reveal the presence of Keplerian disks in only two protostars,
with disk radii of 90 and 200~au. Taken at face value, this suggests
that Keplerian disks larger than \rev{50~au} (the typical scale probed
by our observations) are present around 10-20\% of the Class 0
protostars only. On the other hand, we cannot exclude that some of
these protostars harbor disks with radii larger than \rev{50~au} that
remain undetected in our observations. \rev{For example, Keplerian
  rotation is not detected in L1527 with the \ceoto{} line at
  $r > 74$~au, but we detect Keplerian rotation at $r < 90$~au with
  the \sofs{} line. We argue that this is because the \ceoto{} line
  emission is dominated by the envelope and it is optically thick,
  whereas the \sofs{} is less affected by the envelope and it is
  optically thin. In some other sources, the \ceoto{} may also be too
  optically thick to reveal the Keplerian disk, as it is dominated by
  infalling gas. Unless the Keplerian disk is detected with another
  tracer, the disk would remain undetected. Unfortunately, in most
  sources the \sofs{} line is dominated by the outflow emission, while
  in many others it is not detected on-source above our detection
  limit. Therefore, it cannot be used to constrain the envelope and
  disk kinematics. For this reason, we cannot exclude that more
  sources of our sample harbor a disk larger than 50~au.}

\section{Conclusions}
\label{sec:conclusions} 

We presented Plateau de Bure observations of the
\rev{\tco{}~($J=2-1$), \ceo{}~($J=2-1$) and SO~($N_{j}=5_{6}-4_{5}$)}
line emission in a sample of 16 Class 0 protostars at sub-arcsecond
resolution. These observations were used to constrain the rotation of
the protostars on scales between 50 and 500~au and to search for
Keplerian disks. Our conclusions are the following:

\begin{itemize}

\item Seven Class 0 sources out of 16 (SerpS-MM18, L1448-C, L1448-NB,
  L1527, NGC1333-IRAS2A, NGC1333-IRAS4B, and SVS13-B) show a velocity
  gradient oriented along the direction perpendicular to the jet axis
  (within $\pm 45\degr{}$) in at least one line at a few hundred au
  scales. In other sources, the velocity gradients at these scales are
  mostly due to outflow emission.

\item Among the sources with a gradient approximately orthogonal to
  the jet axis, we detect Keplerian rotation in only two protostars:
  L1527 and L1448-C. Both sources are among the sources of the sample
  with the largest specific angular momentum at 100~au. They are also
  associated with compact continuum emission.

\item In L1527, we detect Keplerian rotation up to a radius of
  0.7\arcsec{} from the continuum peak ($\sim 90$~au) with the \sofs{}
  line. This is larger than the radius derived by
  \citet[][74~au]{2017ApJ...849...56A} from the analysis of the
  \ceoto{} rotation profile, but consistent with the radius of the
  centrifugal barrier obtained by \citet{Sakai14,2017MNRAS.467L..76S}
  from several other molecular tracers. We argue that the difference
  between the disk radius derived with \ceoto{} and \sofs{} lines
  could be explained by the \ceoto{} opacity, but a detailed radiative
  transfer model would be needed to confirm this scenario.
  
\item In L1448-C, we detect Keplerian rotation with the \tcoto{} and
  \ceoto{} lines up to radii of 0.7\arcsec{} ($\sim 200$~au) from the
  continuum peak. This is the first detection of Keplerian rotation in
  this source.

\item On the contrary, we do not detect Keplerian rotation in
  L1448-NB. Our observations and analysis of the \ceoto{} emission
  shows that the velocity profile is proportional to $r^{-1}$ down to
  0.5\arcsec{} (175~au) from the continuum peak.

\item The detection rate of Keplerian disks in the CALYPSO sample of
  Class 0 objects is 10\%. However, we argue that the \tcoto{} and
  \ceoto{} emission could be optically thick in some of the sources as
  it remains dominated by the envelope emission. Unless the disk is
  seen with the \sofs{} line, some of the disks could remain
  undetected by our observations.

\end{itemize}

This study shows that large ($r >50$~au) Keplerian disks are rare in
Class 0 protostars. However, the presence of jets or outflows
\rev{(Podio et al. in prep.) and disk-like dust continuum components
  \citep{2019A&A...621A..76M}} strongly suggests that Class 0
protostars have disks. Therefore, most of these disks are smaller than
50~au, which is consistent with MHD simulations that predict radii on
the order of a few tens of au
\rev{\citep{2016ApJ...830L...8H}}. Finding and characterizing
Keplerian disks around Class 0 protostars will require a new survey of
a sample of Class 0 protostars with a spatial resolution of
0.1\arcsec{} or better. This can only be achieved by ALMA.

\begin{acknowledgements}

  We would like to thank Nagayoshi Ohashi, Nami Sakai and John Tobin
  for fruitful discussions on the interpretation of the observations
  presented in this paper. \rev{We also thank the referee for his/her
    detailed report that helped us to improve the paper}. This work is
  based on observations carried out under project number U052 with the
  IRAM NOEMA Interferometer. IRAM is supported by INSU/CNRS (France),
  MPG (Germany) and IGN (Spain). This work has benefited from the
  support of the European Research Council under the European Union’s
  Seventh Framework Programme (Advanced Grant ORISTARS with grant
  agreement No. 291294 and Starting Grant MagneticYSOs with grant
  agreement No. 679937), and from the French Agence Nationale de la
  Recherche (ANR), under reference ANR-12-JS05-0005.

\end{acknowledgements}

\bibliographystyle{aa}
\bibliography{biblio}

\begin{appendix}

  \section{Synthesized beam sizes and sensitivities}
\label{sec:synth-beam-sens}

Synthesized beam sizes and sensitivities for each line are given in
Tables~\ref{tab:13co},~\ref{tab:c18o}, and \ref{tab:so}.

\begin{table}
  \begin{center}
    \caption{Beam sizes and noise per channel for the \tcoto{} line
      observations. \label{tab:13co}}
    \begin{tabular}{lllll}
      \hline
      \hline
      Source & \multicolumn{3}{c}{Synthesized beam} & Noise                                       \\
      \cline{2-4}
             & Major                                & Minor       & PA        & (\mjy{}         \\
             & ($\arcsec{}$)                        & ($\arcsec$) & ($\degr{}$) & \, beam$^{-1}$) \\
      \hline
      L1448-2A       & 0.63 & 0.41 & 33   & 14 \\
      L1448-NB       & 0.80 & 0.76 & 50   & 19 \\
      L1448-C        & 0.65 & 0.41 & 32   & 13 \\
      NGC1333-IRAS2A & 0.81 & 0.77 & 47   & 18 \\
      SVS13B         & 0.74 & 0.62 & 37   & 12 \\
      NGC1333-IRAS4A & 0.77 & 0.64 & 37   & 14 \\
      NGC1333-IRAS4B & 0.80 & 0.68 & 28   & 14 \\
      IRAM04191      & 0.69 & 0.47 & 33   & 15 \\
      L1521F         & 0.63 & 0.41 & 33   & 11 \\
      L1527          & 0.72 & 0.64 & 48   & 11 \\
      SerpM-S68N     & 1.11 & 0.56 & -156 & 24 \\
      SerpM-SMM4     & 1.16 & 0.68 & 29   & 22 \\
      SerpS-MM18     & 1.17 & 0.66 & 22   & 23 \\
      SerpS-MM22     & 1.25 & 0.59 & -161 & 19 \\
      L1157          & 0.63 & 0.51 & -178 & 19 \\
      GF9-2          & 0.53 & 0.37 & 5    & 11 \\
      \hline
    \end{tabular}
  \end{center}
\end{table}

\begin{table}
  \begin{center}
    \caption{Beam sizes and noise per channel for the \ceoto{} line
      observations. \label{tab:c18o}}
    \begin{tabular}{lllll}
      \hline
      \hline
      Source & \multicolumn{3}{c}{Synthesized beam} & Noise                                       \\
      \cline{2-4}
             & Major                                & Minor       & PA        & (\mjy{}         \\
             & ($\arcsec{}$)                        & ($\arcsec$) & ($\degr{}$) & \, beam$^{-1}$) \\
      \hline
      L1448-2A       & 0.63 & 0.41 & 33   & 13 \\
      L1448-NB       & 0.81 & 0.76 & 38   & 18 \\
      L1448-C        & 0.65 & 0.41 & 32   & 13 \\
      NGC1333-IRAS2A & 0.81 & 0.77 & 44   & 15 \\
      SVS13B         & 0.74 & 0.62 & 37   & 11 \\
      NGC1333-IRAS4A & 0.77 & 0.64 & 34   & 14 \\
      NGC1333-IRAS4B & 0.80 & 0.68 & 28   & 14 \\
      IRAM04191      & 0.69 & 0.47 & 33   & 14 \\
      L1521F         & 0.63 & 0.41 & 33   & 10 \\
      L1527          & 0.72 & 0.64 & 46   & 11 \\
      SerpM-S68N     & 1.12 & 0.57 & -156 & 20 \\
      SerpM-SMM4     & 1.14 & 0.67 & 30   & 21 \\
      SerpS-MM18     & 1.17 & 0.66 & 22   & 22 \\
      SerpS-MM22     & 1.26 & 0.60 & -161 & 18 \\
      L1157          & 0.64 & 0.51 & -177 & 18 \\
      GF9-2          & 0.53 & 0.37 & 5    & 10 \\
      \hline
    \end{tabular}
  \end{center}
\end{table}

\begin{table}
  \begin{center}
    \caption{Beam sizes and noise per channel for the \sofs{} line
      observations. \label{tab:so}}
    \begin{tabular}{lllll}
      \hline
      \hline
      Source & \multicolumn{3}{c}{Synthesized beam} & Noise                                       \\
      \cline{2-4}
             & Major                                & Minor       & PA        & (\mjy{}         \\
             & ($\arcsec{}$)                        & ($\arcsec$) & ($\degr{}$) & \, beam$^{-1}$) \\
      \hline
      L1448-2A       & 0.63 & 0.41 & 33   & 13 \\
      L1448-NB        & 0.81 & 0.76 & 39   & 18 \\
      L1448-C        & 0.65 & 0.41 & 32   & 13 \\
      NGC1333-IRAS2A & 0.80 & 0.77 & 69   & 16 \\
      SVS13B         & 0.74 & 0.62 & 37   & 11 \\
      NGC1333-IRAS4A & 0.77 & 0.64 & 34   & 15 \\
      NGC1333-IRAS4B & 0.79 & 0.69 & 15   & 14 \\
      IRAM04191      & 0.69 & 0.47 & 30   & 14 \\
      L1521F         & 0.63 & 0.41 & 33   & 10 \\
      L1527          & 0.72 & 0.64 & 48   & 11 \\
      SerpM-S68N     & 1.12 & 0.57 & -156 & 24 \\
      SerpM-SMM4     & 1.14 & 0.67 & 30   & 21 \\
      SerpS-MM18     & 1.17 & 0.66 & 22   & 24 \\
      SerpS-MM22     & 1.34 & 0.61 & -159 & 18 \\
      L1157          & 0.64 & 0.51 & -177 & 18 \\
      GF9-2          & 0.53 & 0.37 & 5    & 12 \\
      \hline
    \end{tabular}
  \end{center}
\end{table}

  \section{Moment maps}
\label{sec:moment-maps}

\begin{figure*}
  \includegraphics[width=\hsize]{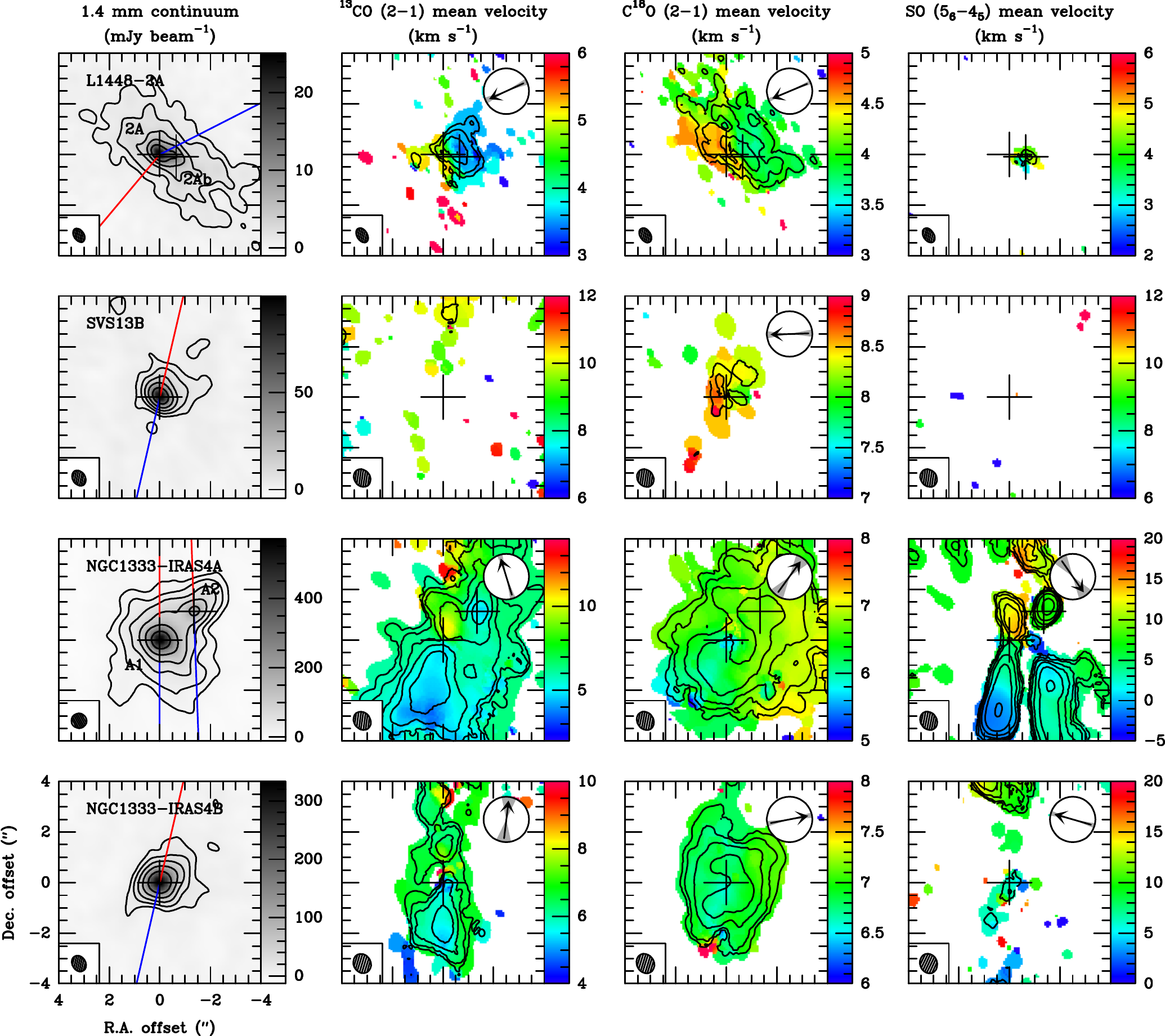}
  \caption{Same as in Fig.~\ref{momentmaps-1} for L1448-2A, SVS13B
    NGC1333-IRAS4A and NGC1333-IRAS4B. \label{momentmaps-2}}
\end{figure*}

\begin{figure*}
  \includegraphics[width=\hsize]{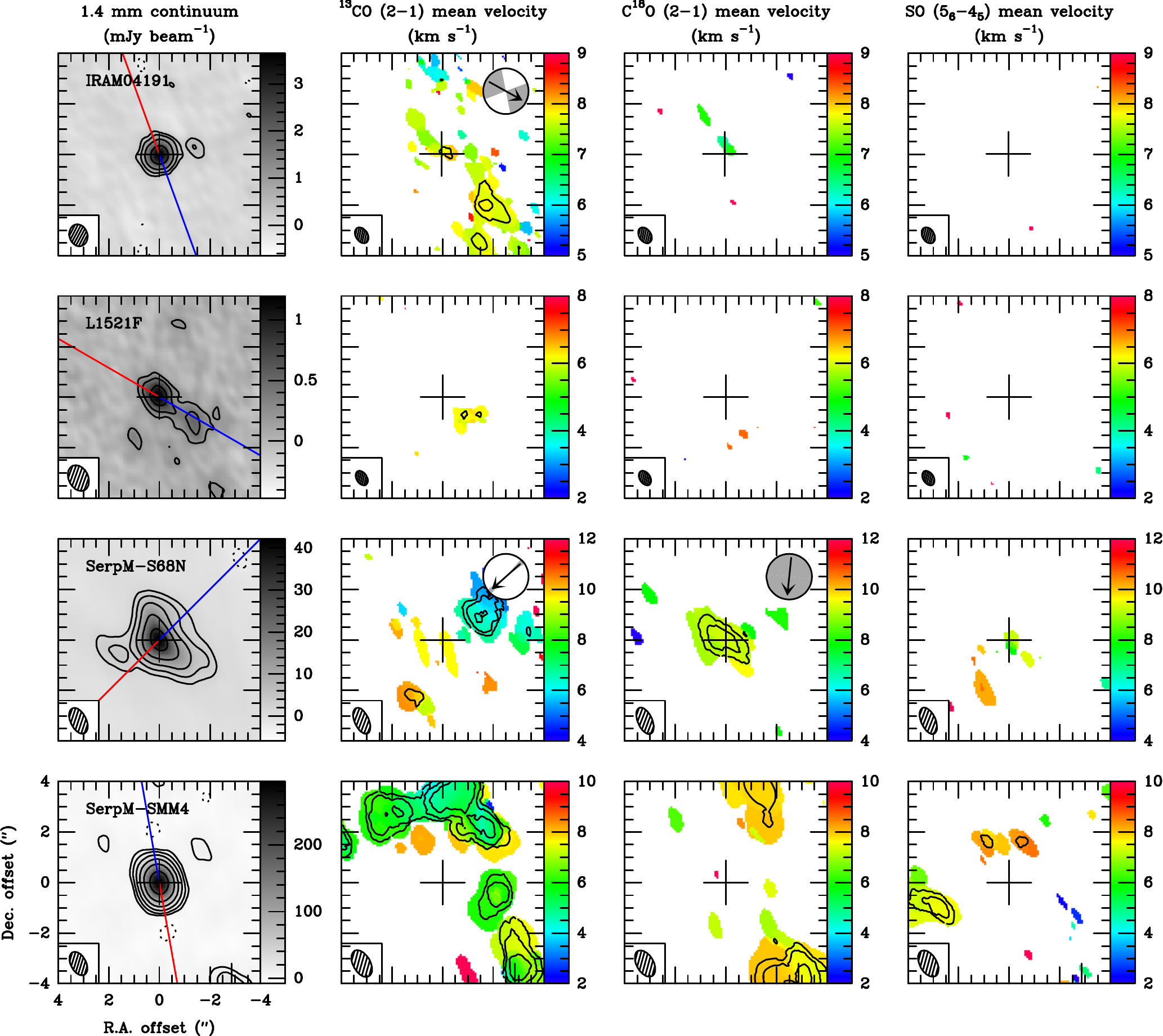}
  \caption{Same as in Fig.~\ref{momentmaps-1} for IRAM04191, L1521F,
    SerpM-S68N, and SerpM-SMM4. \label{momentmaps-3}}
\end{figure*}

\begin{figure*}
  \includegraphics[width=\hsize]{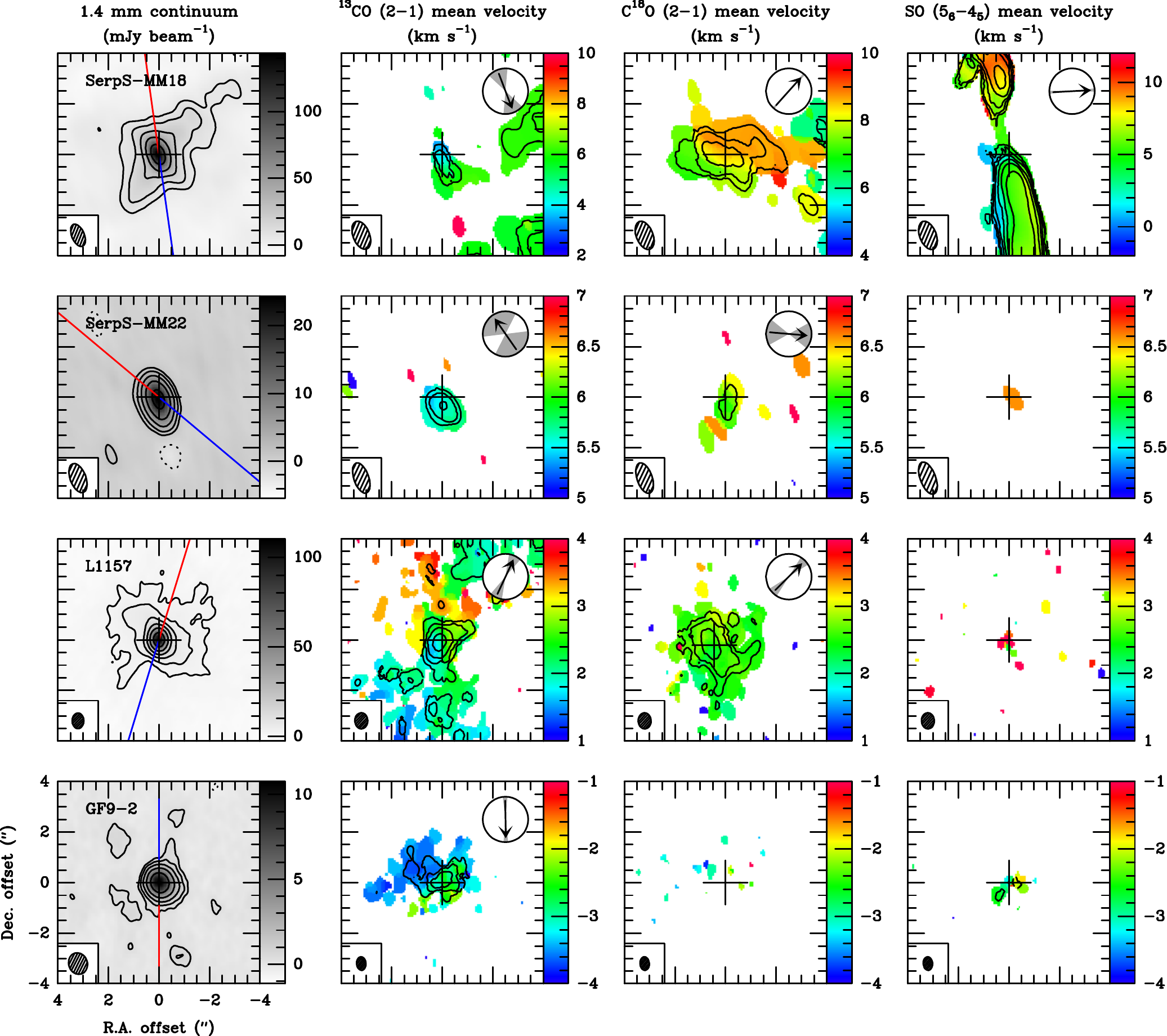}
  \caption{Same as in Fig.~\ref{momentmaps-1} for SerpS-MM18,
    SerpS-MM22, L1157, and GF9-2. \label{momentmaps-4}}
\end{figure*}

Figures~\ref{momentmaps-2}, \ref{momentmaps-3}, and \ref{momentmaps-4}
show zeroth-order moment and first-order moment (mean velocity) maps
of the \tcoto{}, \ceoto{} and \sofs{} lines, together with the 1.4 mm
continuum emission maps from Maury et al. (2019) in L1448-2A, SVS13B,
NGC1333-IRAS4A, NGC1333-IRAS4B, IRAM04191, L1521F, SerpM-S68N,
SerpM-SMM4, SerpS-MM18, SerpS-MM22, L1157, and GF9-2.

  \section{Other disk candidates}
\label{sec:other-disk-cand}

\subsection*{NGC1333-IRAS2A}
\label{sec:ngc1333-iras2a}

NGC1333-IRAS2A (hereafter IRAS2A) is a Class 0 protostar located in
the NGC1333 region, within the Perseus cloud. It is the most luminous
source of our sample, with an internal luminosity of 47~L$_\odot$
(Ladejate et al. in prep.). Its envelope mass is 7.9~M$_\odot$
\citep{2013A&A...552A.141K}. The PA of the main jet is -155\degr{}
(Podio et al. in prep.). A second jet originates from a few arcseconds
south of IRAS2A, and has a PA of -65\degr{} \citep[][Podio et al. in
prep.]{Codella14}. \citet{2009A&A...502..199B} studied the kinematics
of the IRAS2A envelope using interferometric observations at
1\arcsec{} resolution and find that the gas kinematics is dominated by
infall, with very little rotation. Using CH$_{3}$OH line observations
with a 0.8\arcsec{} resolution, \citet{Maret14} measured a small
velocity gradient orthogonal to the jet axis (PA 107\degr{}).  They
argue that the methanol emission originates from the infalling and
slowly rotating envelope around a central mass of 0.1 - 0.2~M$_\odot$.

First-order moment maps for the \tcoto{}, \ceoto{}, and \sofs{} lines
are shown in Fig.~\ref{momentmaps-1}. The \tcoto{} emission peaks at
the same position as the continuum peak. The mean velocity at this
position is 4.5~\kms{}, \rev{blueshifted from} the source systemic
velocity ($\sim$~7.5~\kms{})\footnote{The P.V. diagram for the
  \ceoto{} emission, which appears to trace the rotation in the
  envelope, is peaked around $\vlsr = 7.5$~\kms{} (see
  Fig.~\ref{fig:pv-iras2a}). We therefore adopt a systemic velocity of
  7.5~\kms{} for this source.}. This suggests that the \tcoto{}
emission is optically thick and possibly in part filtered out by the
interferometer. The mean velocity increases to $\sim$12~\kms{} toward
the north-east, in the direction of the red-shifted lobe of the
jet. West of the continuum peak, the mean velocity is comparable to
the systemic velocity. The \ceoto{} line also peaks close to the
continuum peak and shows a velocity gradient along an east-west
axis. The \sofs{} line indicates a velocity gradient along a
north-south axis and appears to probe the outflow. Fitting the first
order moment maps, we measure a gradient PA of $(-29 \pm 12)$~\degr{},
$(60 \pm 4)$~\degr{} and $(-1 \pm 6)$~\degr{} for the \tcoto{},
\ceoto{} and \sofs{} lines, respectively. The corresponding values of
$\Delta \theta$ are 36\degr{}, 55\degr{}, and 64\degr{},
respectively. Only the \tcoto{} line fulfills the
$\Delta \theta < 45$~\degr{} criterion, and it could, in principle,
probe the inner envelope or the disk. However, it is unlikely to be
the case because the orientation of the gradient we measure with this
line (PA = $(-29 \pm 12)$\degr{}) is inconsistent with that measured
by \cite{Maret14} from CH$_{3}$OH line observations (PA = 107\degr{}).

\begin{figure*}
  \includegraphics[width=\hsize]{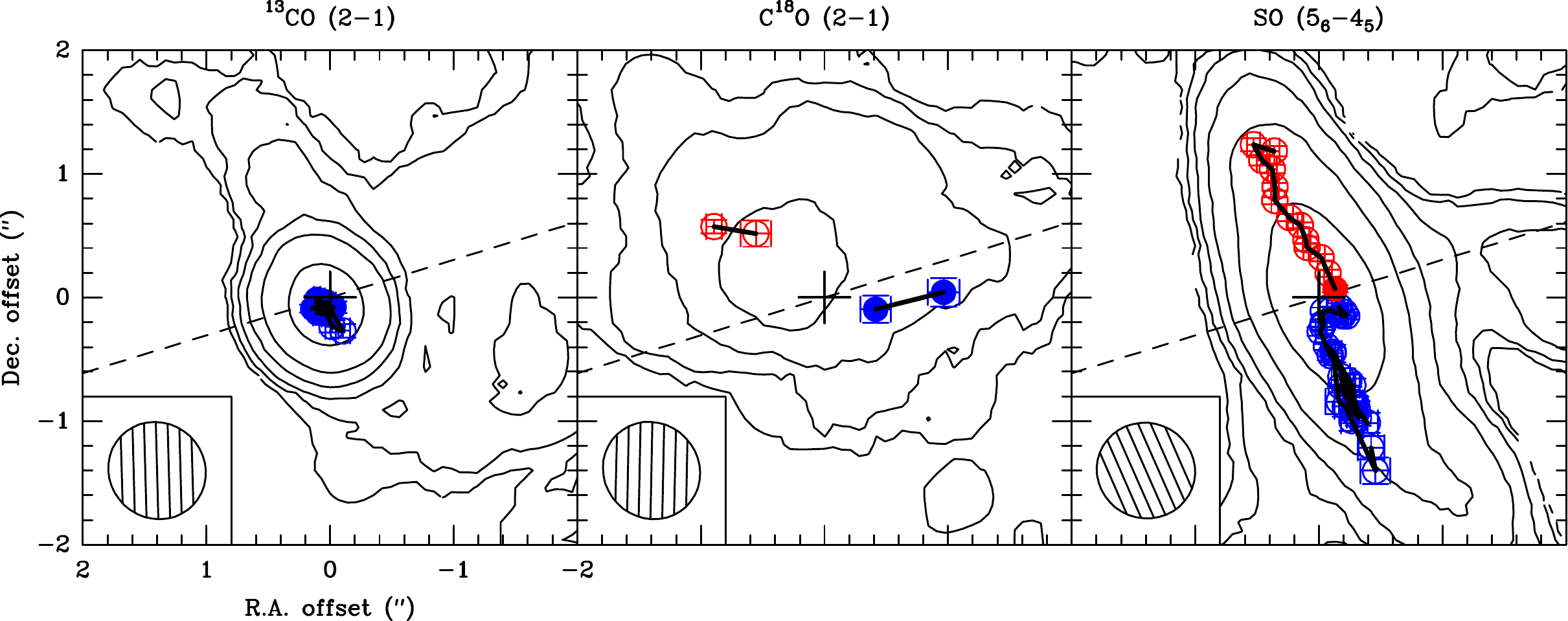}
  \caption{Same as in Fig.~\ref{fig:uvfit-l1527} for
    IRAS2A. \label{fig:uvfit-iras2a}}
\end{figure*}

\begin{figure*}
  \includegraphics[width=\hsize]{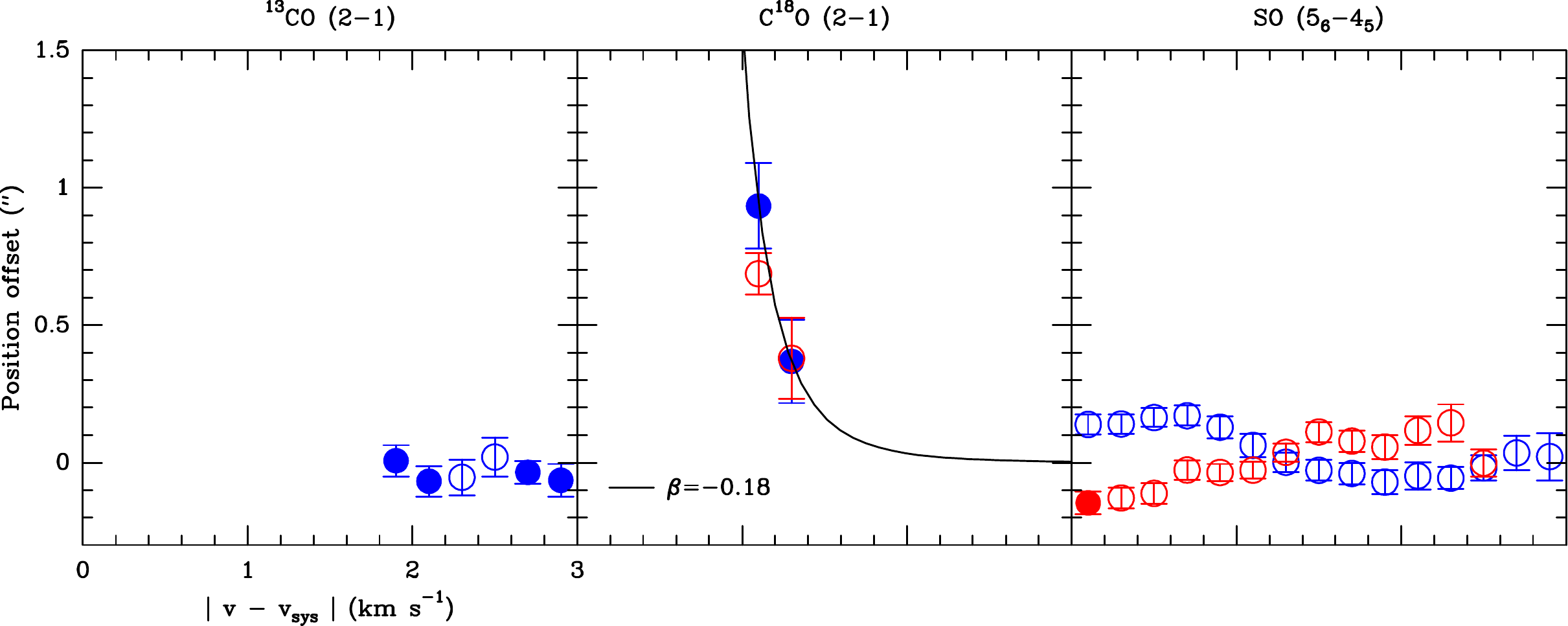}
  \caption{Same as in Fig.~\ref{fig:velocityfit-l1527} for
    IRAS2A. \label{fig:velocityfit-iras2a}}
\end{figure*}

\begin{figure*}
  \includegraphics[width=\hsize]{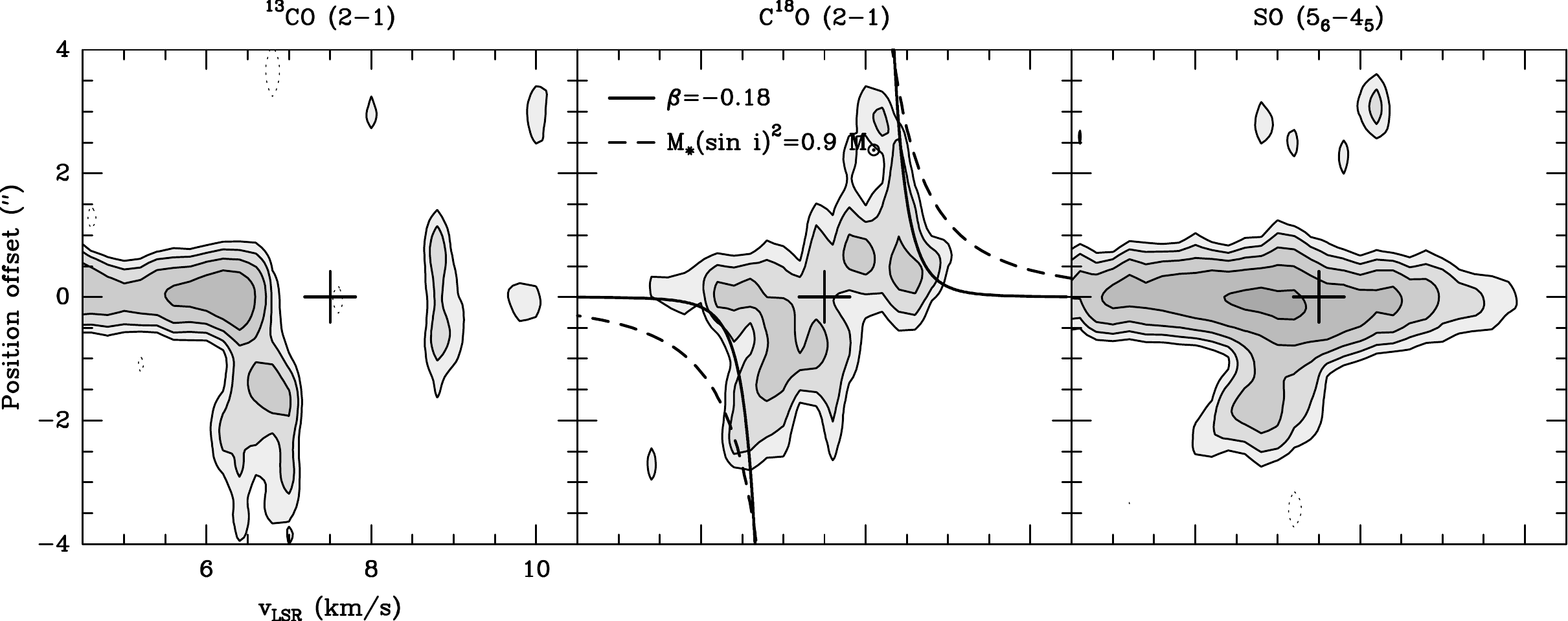}
  \caption{Same as in Fig.~\ref{fig:pv-l1527} for IRAS2A. The dashed
    curve in the middle panel shows the Keplerian velocity for
    $\Msinisqr = 0.9~\Msun$. \label{fig:pv-iras2a}}
\end{figure*}

Figure~\ref{fig:uvfit-iras2a} shows the centroid positions together
with the zeroth-order moment maps. For consistency with
\citet{Maret14}, we assume that the disk major axis has a PA =
107\degr{}, slightly different from the direction orthogonal to the
main jet (PA = 115\degr{}). For the \tcoto{} emission, several
blue-shifted channels can be fitted, and their centroid positions are
close to the continuum peak. These channels correspond to large
velocities ($|v - v_\mathrm{sys}| > 2$~\kms{}), so it is possible that
they are due to \rev{outflow emission}. For the \sofs{} emission, the
centroids are located along the jet axis. For the \ceoto{} line, we
can fit the position of four channels. The centroids of the two
blue-shifted channels are roughly consistent with the gradient
orientation measured by \citet{Maret14}. However, the centroids of the
two red-shifted channels are not aligned with the gradient
orientation. The centroids of the \sofs{} line are aligned with the
jet. The corresponding rotation curves are shown in
Fig.~\ref{fig:velocityfit-iras2a}. We do not attempt to fit the data
points for the \tcoto{} and \sofs{} line, because the emission of
these two lines is most likely due to outflow emission. For the
\ceoto{} line, we obtain a best-fit $\beta = -0.18 \pm 0.09$.  However
the fit is uncertain since it is based on two channels only. In
Fig.~\ref{fig:pv-iras2a}, we show PV diagrams obtained along the same
axis. The \ceoto{} emission diagram is consistent with infall and
rotation along this axis, in agreement with the results of
\citet{Maret14}. Fitting the first emission contour with a Keplerian
law gives $\Msinisqr = 0.9~\Msun$. The diagrams for the \rev{\tcoto{}}
and \sofs{} lines indicates the presence high-velocity gas close to
the central object, which is likely due to \rev{the outflow}.

In summary, we find evidence for rotation and infall in this source
from the \ceoto{} line observations. The rotation profile we derive
with this line is uncertain, but it does not appear to be consistent
with Keplerian rotation.

\subsection*{SVS13B}
\label{sec:svs13b}

SVS13B is a Class 0 protostar also located in NGC1333. This source is
part of a wide triple system \citep{2017A&A...604L...1L}. SVS13A is a
Class~I protostar located 4400~au (15\arcsec{}) from SVS13B, while
SVS13C, a Class 0 protostar, is located 5600~au (19\arcsec) from
SVS13B \citep{Tobin16a,2019A&A...621A..76M}. The internal luminosity
of SVS13B is $3 \pm 2$~L$_{\odot}$ (Ladejate et al. in prep.)  and its
envelope mass is 1.9~M$_{\odot}$ \citep{1997A&A...325..542C}. It
drives a collimated outflow seen in high velocity SiO
\citep{1998A&A...339L..49B}, with a PA of 167\degr{} (Podio et al. in
prep.). Based on VLA observations and modeling, \cite{SeguraCox16}
argue that SVS13B harbors a disk with a radius of $\sim 40$~au
(0.13\arcsec{}).

Our observations of the \tcoto{}, \ceoto{} and \sofs{} line emission
in SVS13B are shown in Fig.~\ref{momentmaps-2}. Only \ceoto{} emission
is detected towards the continuum peak. The emission is compact, with
a size of about 2\arcsec{}. The first-order moment map shows a
velocity gradient, with the mean velocity increasing from 8~\kms{}
west of the continuum peak up to 9~\kms{} east of it. Fitting the
first-order moment map, we find a gradient position angle
$\theta = (92 \pm 9)\degr{}$, close to the disk orientation expected
from the jet axis ($\Delta \theta = 15\degr{}$). The gradient
orientation is roughly consistent with the disk PA derived by
\citet[][PA = 71\degr{}]{SeguraCox16}.

\begin{figure}
  \centering \includegraphics[width=7.0cm]{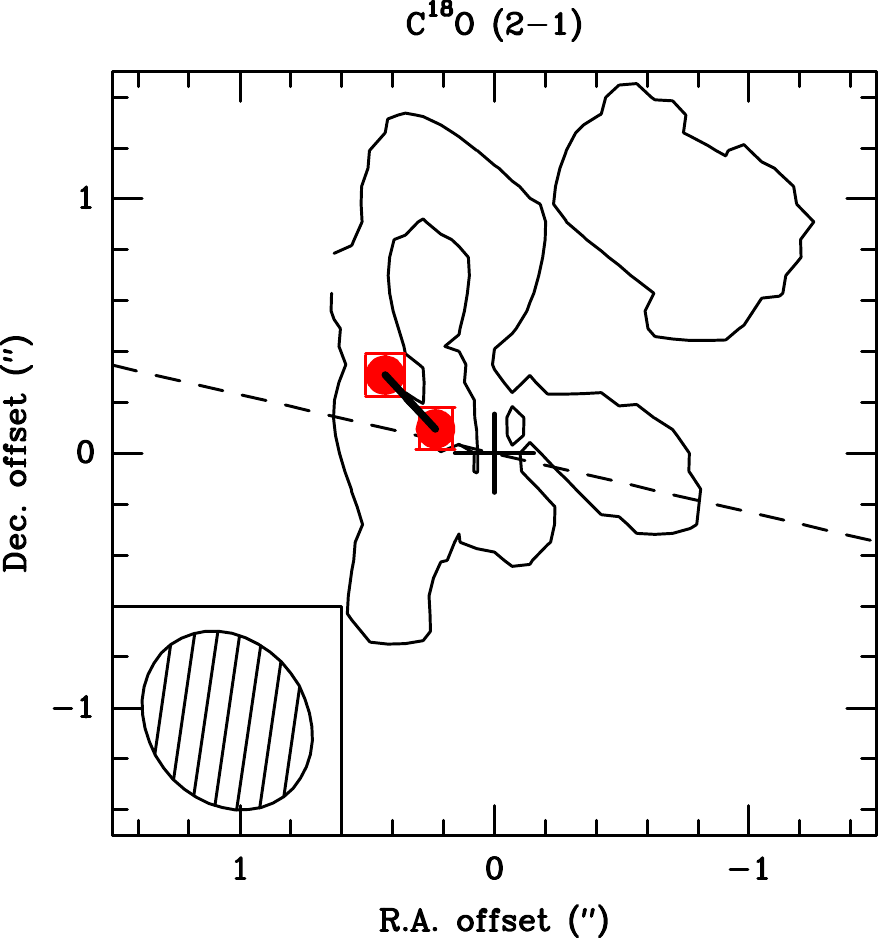}
  \caption{Same as in Fig.~\ref{fig:uvfit-l1527} for
    SVS13B. \label{fig:uvfit-svs13b}}
\end{figure}

\begin{figure}
  \centering \includegraphics[width=7.0cm]{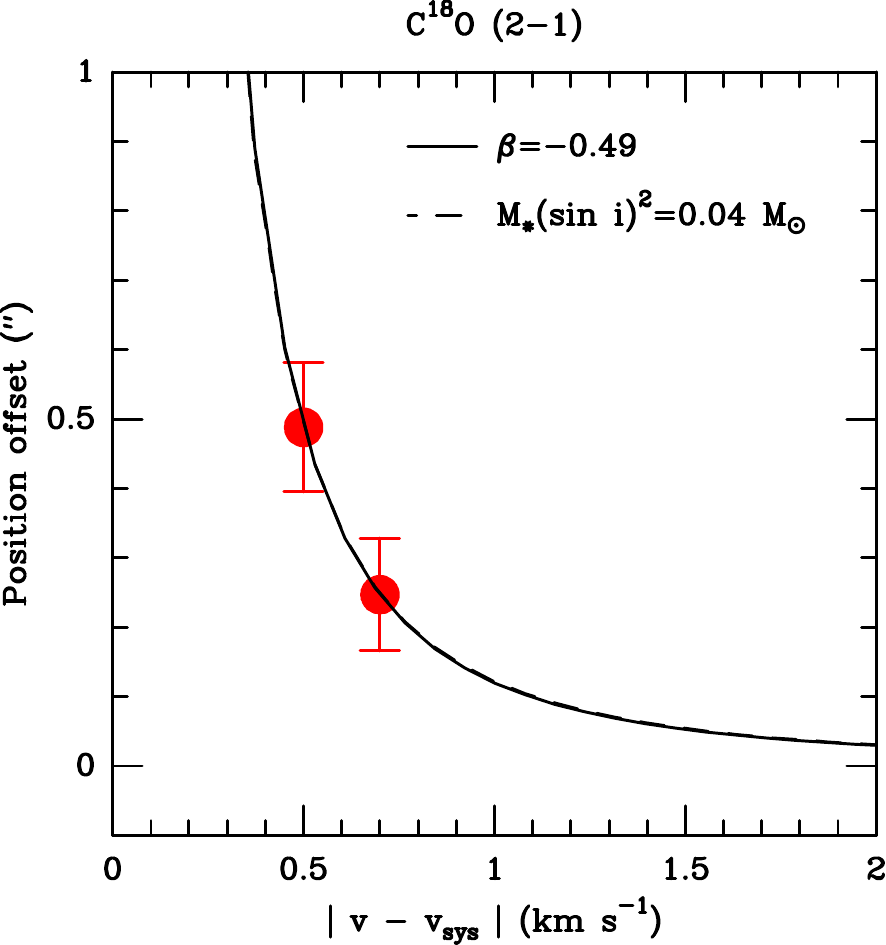}
  \caption{Same as in Fig.~\ref{fig:velocityfit-l1527} for
    SVS13B. \label{fig:velocityfit-svs13b}}
\end{figure}

Figure~\ref{fig:uvfit-svs13b} shows the positions of the centroids we
obtain from a fit in the $uv$ plane, together with the zeroth-order
moment. Only two red-shifted channels can be fitted in the $uv$
plane. The centroids of both channels are located east of the
continuum peak, in agreement with the velocity gradient orientation we
measure in this source. The rotation curve we obtain is shown in
Fig.~\ref{fig:velocityfit-svs13b}. For this source, we adopt a
systemic velocity of \rev{8.4~\kms{} \citep{2009ApJ...691.1729C}} and
we assume that the disk is orthogonal to the jet axis (PA 77
\degr{}). From a fit with a power law, we obtain a best-fit
\rev{$\beta = -0.49 \pm 0.28$} for $r < 0.5\arcsec$ (150~au),
consistent with Keplerian rotation. However, we caution that the
rotation curve for this source is very uncertain, because it is based
on two velocity channels only. The lowest velocity channel we fit is a
position offset of 0.5\arcsec, i.e.\rev{,} a radius of 150~au. From a
fit with a Keplerian law, we find a best-fit
\rev{$\Msinisqr = 0.04 \pm 0.01$~M$_\odot{}$}.

\begin{figure*}
  \sidecaption
  \includegraphics[width=12cm]{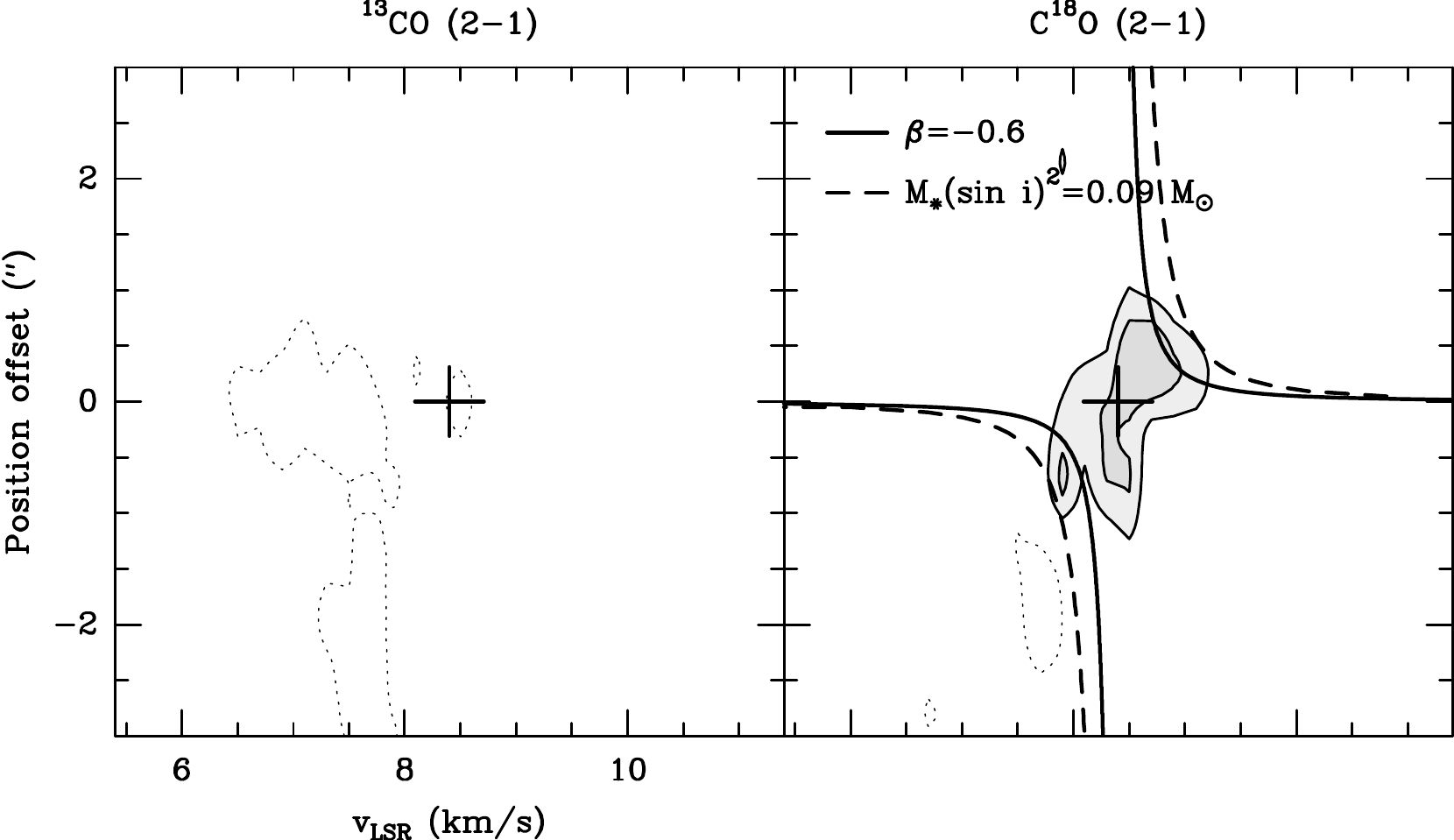}
  \caption{Same as in Fig.~\ref{fig:pv-l1527} for SVS13B. Dashed curve
    in the right panel shows the Keplerian velocity for
    $\Msinisqr = 0.09~\Msun$\label{fig:pv-svs13b}}
\end{figure*}

Figure~\ref{fig:pv-svs13b} shows the PV diagram we obtain along the
assumed disk axis, together with the best-fit power-law rotation
curve. The PV diagram appears to be symmetric around the adopted value
of $v_\mathrm{sys}$ and it is consistent with rotation. However, the
emission is at the limit of the spatial resolution of our
observations. Fitting the first emission contour with a Keplerian law
gives $\Msinisqr = 0.09 \pm 0.01$~M$_\odot{}$.

To summarize, we tentatively detect Keplerian rotation at $r < 150$~au
in this source. From a fit in the $uv$ plane and the PV diagram, we
estimate that \rev{$\Msinisqr = 0.04 - 0.09$~\Msun{}}, that is\rev{,}
\rev{$\Mstar = 0.05 - 0.12$~\Msun{}}, assuming a disk inclination of
61~\degr{} \citep{SeguraCox16}. Observations at higher angular
resolution are needed to confirm this detection.

\subsection*{NGC1333-IRAS4B}
\label{sec:ngc1333-iras4b}

NGC1333-IRAS4B (hereafter IRAS4B) is a Class 0 protostar also located
in NGC1333. This source is a wide binary whose secondary component,
IRAS4B2, is located 11\arcsec{} (3200~au) to the east of the main
component \citep[]{2007ApJ...659..479J,2019A&A...621A..76M}. The
internal luminosity of IRAS4B is 2.3~L$_\odot$ (Ladejate et al. in
prep.), and it is surrounded by a 4.7~M$_\odot$ envelope
\citep{Sadavoy14}. IRAS4B drives a jet oriented along a north-south
axis (PA = 167\degr{}; Podio et al. in prep.). Using \ceoto{}
observations, \citet{Yen13,Yen15a} measured a velocity gradient with a
PA of -14\degr{}, which is dominated by the outflow. Fitting only
the component orthogonal to the outflow, they find little rotation of
the envelope and derive a centrifugal radius $< 5$~au.

Our observations of IRAS4B are shown in Fig.~\ref{momentmaps-2}. The
\tcoto{} emission is elongated along the jet direction, and shows a
velocity gradient that is consistent with the jet orientation. \sofs{}
emission is detected close to the continuum peak, as well as in a few
spots along the jet axis. The \ceoto{} emission is centered close to
the continuum emission, and its size is $\sim$~6\arcsec{} in the
north-south direction, and $\sim$~4\arcsec{} in the east-west
direction. The \ceoto{} first-order moment map hints at a small
velocity gradient along the east-west direction: the mean velocity to
the east of the source is about 6.5~\kms{}, while it is 7.0~\kms{}
west of it. This gradient is confirmed by the fit of the first-order
moment map, which gives a PA of $-79 \pm 6$\degr{} and
$\Delta \theta = 24$~\degr{}. For the \tcoto{} line, we find a
velocity gradient with $\theta = (-7 \pm 19)$\degr{} and
$\Delta \theta = 84$~\degr{}. This confirms that the line is mostly
due to outflow emission. For the \sofs{}, we find
$\theta = (74 \pm 6)$\degr{} and $\Delta \theta = 3$~\degr{}. Although
the orientation of the gradient we derive for this line is almost
orthogonal to the jet, it is unlikely to be due to envelope rotation
given the morphology of the \sofs{} emission on larger scales. In
addition, the orientation of the gradient measured with the \sofs{}
line is in a direction opposite to the \ceoto{} line velocity
gradient.

\begin{figure*}
  \includegraphics[width=\hsize]{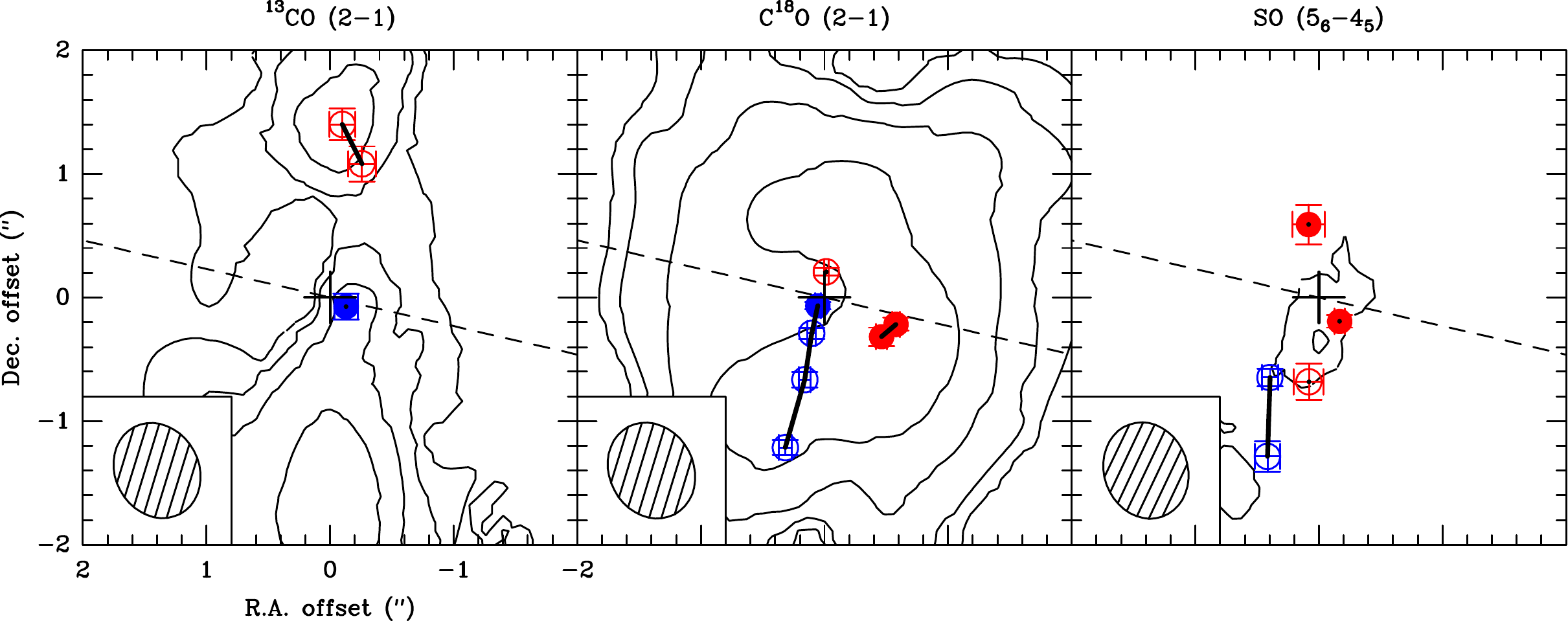}
  \caption{Same as in Fig.~\ref{fig:uvfit-l1527} for
    IRAS4B. \label{fig:uvfit-iras4b}}
\end{figure*}

\begin{figure*}
  \includegraphics[width=\hsize]{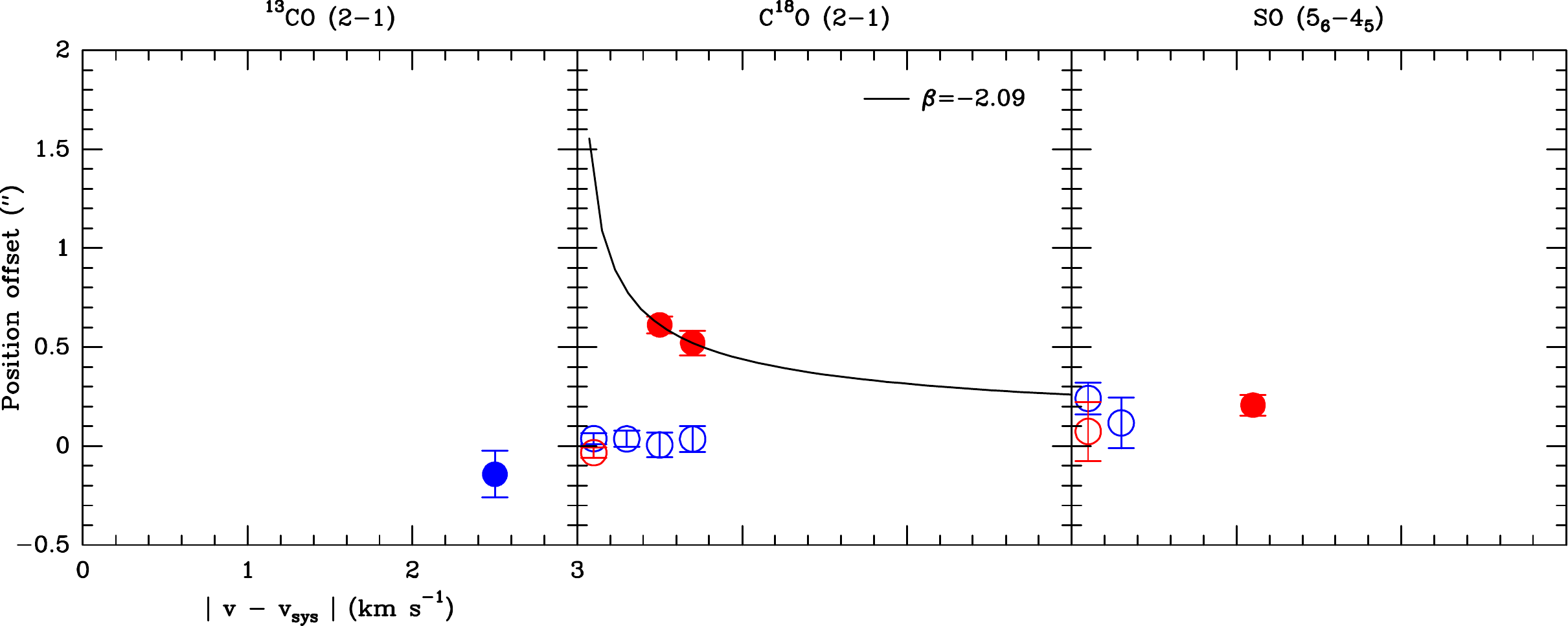}
  \caption{Same as in Fig.~\ref{fig:velocityfit-l1527} for
    IRAS4B. \label{fig:velocityfit-iras4b}.}
\end{figure*}

Figure~\ref{fig:uvfit-iras4b} shows the results of the fit in the $uv$
plane. We assume that the disk major axis is orthogonal to the jet (PA
-103\degr{}), and we adopt a systemic velocity of 6.7~\kms{},
consistent with the PV diagram of the \ceoto{} emission (see
below). For both the \tcoto{} and \sofs{} lines, the centroids for the
blue-shifted channels are located to the south, while the centroids
for red-shifted channels are oriented to the north. This is consistent
with the jet orientation. For the \ceoto{} lines, the centroids of the
red-shifted channels are located west of the source, along the
expected disk axis. The centroids for the blue-shifted channels are
roughly aligned along a north-south axis, and are probably
contaminated by the outflow. Figure~\ref{fig:velocityfit-iras4b} shows
the position offset along the assumed disk axis as a function of
velocity. We do not attempt to fit a rotation curve for the \tcoto{}
and \sofs{} lines, because these two lines are dominated by the
outflow. For the \ceoto{} line, we fit a rotation curve considering
only two red-shifted channels. We obtain a best fit
$\beta = -2.09 \pm 1.78$. The fit for this source is uncertain,
because it is based on two points only.

\begin{figure*}
  \includegraphics[width=\hsize]{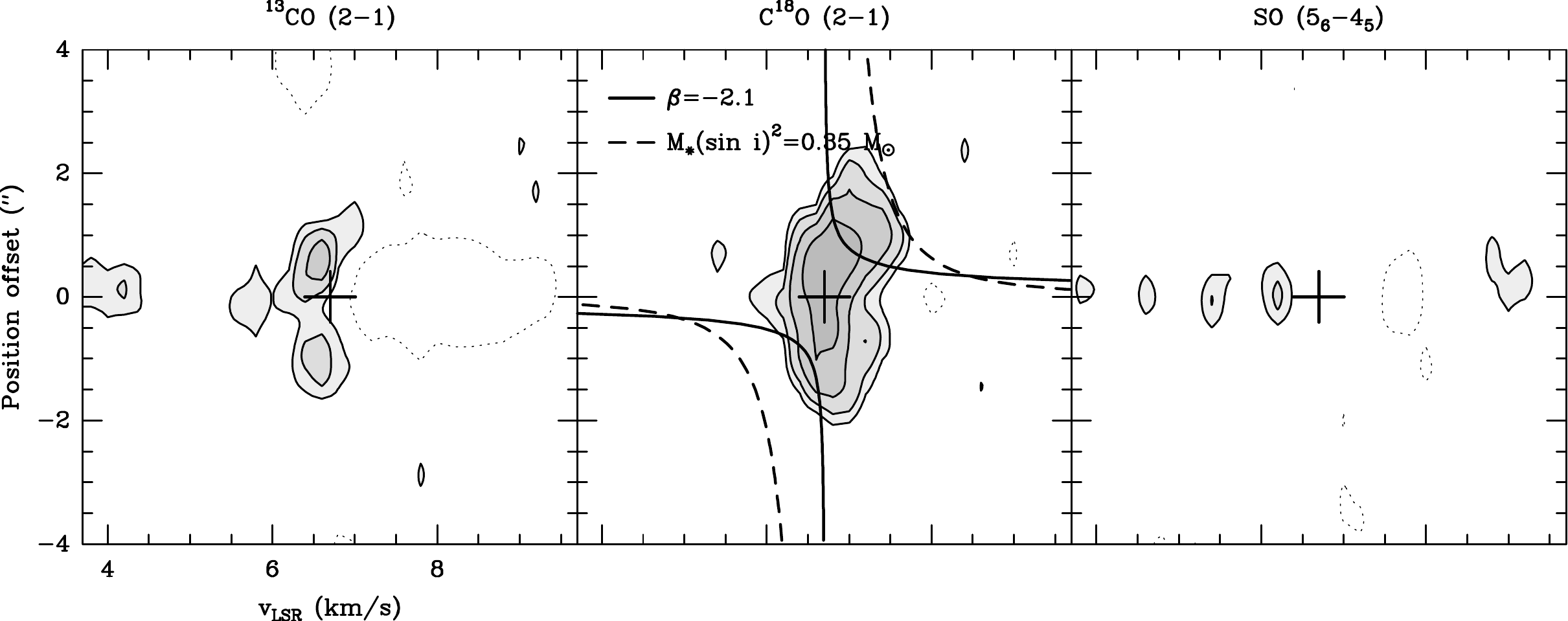}
  \caption{Same as in Fig.~\ref{fig:pv-l1527} for IRAS4B. Dashed curve
    in the middle panel shows the Keplerian velocity for
    $\Msinisqr = 0.35~\Msun$ \label{fig:pv-iras4b}}
\end{figure*}

Figure~\ref{fig:pv-iras4b} shows the PV diagrams along the assumed
disk axis. The PV diagram for the \tcoto{} line shows negative
emission at $\vlsr{} > \vsys{}$. This is likely due to spatial
filtering and/or absorption by a foreground component. Emission is
detected on several spots at $\vlsr{} < \vsys{}$ which are probably
due to shocked gas. The diagram for the \sofs{} shows a similar
pattern. On the other hand, The emission contours for the \ceoto{}
line peak close to a velocity of about 6.7~\kms{}, and appear to be
consistent with envelope rotation. From a fit of the first contour, we
obtain $\Msinisqr = 0.35~\Msun$.

To summarize, we find evidence for rotation from our \ceoto{} line
observations. The velocity curve we derive is uncertain, so we cannot
determine if the rotation is Keplerian.

\subsection*{SerpS-MM18}
\label{sec:serps-mm18}

SerpS-MM18 is a Class 0 protostar located in the Serpens-South region
of the Serpens/Aquila complex, at a distance of 350~pc (Palmeirim et
al., in prep.).  This protostar is part of a wide binary system,
SerpS-MM18 and SerpS-MM18', separated by 3600~au
\citep{2019A&A...621A..76M}. The internal luminosity of SerpS-MM18 is
29~L$_\odot{}$ and its envelope mass is 5~M$_\odot{}$ \cite[Ladejate
et al. in prep.; ][]{2011A&A...535A..77M}. The source drives a jet
with a PA of -172\degr{} (Podio et al in prep.).

Our observations of this source are shown in
Fig.~\ref{momentmaps-4}. We detect \tcoto{} emission towards the
continuum peak position that extends towards the south-west, along the
direction of the blue-shifted lobe of the jet. We also detect \sofs{}
emission along both lobes of the jet. On the contrary, the \ceoto{}
emission appears to originate mostly in the envelope around the
protostar. The \ceoto{} emission peaks at the same position as the
continuum and it is extended along an east-west axis. We observe a
velocity gradient along southeast-northwest axis, with a mean velocity
of 6.5~\kms{} to the southeast of the continuum peak, increasing up to
8.5~\kms{} to the northwest. From a fit of the first-order moment map,
we obtain $\theta = (-159 \pm 26)\degr$, $\theta = (-42 \pm 3)\degr$,
and $\theta = (-87 \pm 3)\degr$ for the \tcoto{}, \ceoto{}, and
\sofs{} lines, respectively. The corresponding values of
$\Delta \theta$ are 77\degr{}, 40\degr{} and 5\degr{},
respectively. The gradient orientation we obtain for \tcoto{} is
consistent with the jet orientation. However, the gradient for the
\sofs{} line is almost perpendicular to jet, although the emission of
this line is clearly dominated by the outflow. Indeed, a careful
examination of the first-order moment map for that line shows that the
eastern part of the blue-shifted jet lobe has a lower mean velocity
than the western part, suggesting rotation of the outflow
cavities. The gradient we measure for the \ceoto{} line is consistent
with envelope or disk rotation, although the value of $\Delta \theta$
we measure is close to the 45$\degr{}$ threshold.

\begin{figure*}
  \includegraphics[width=\hsize]{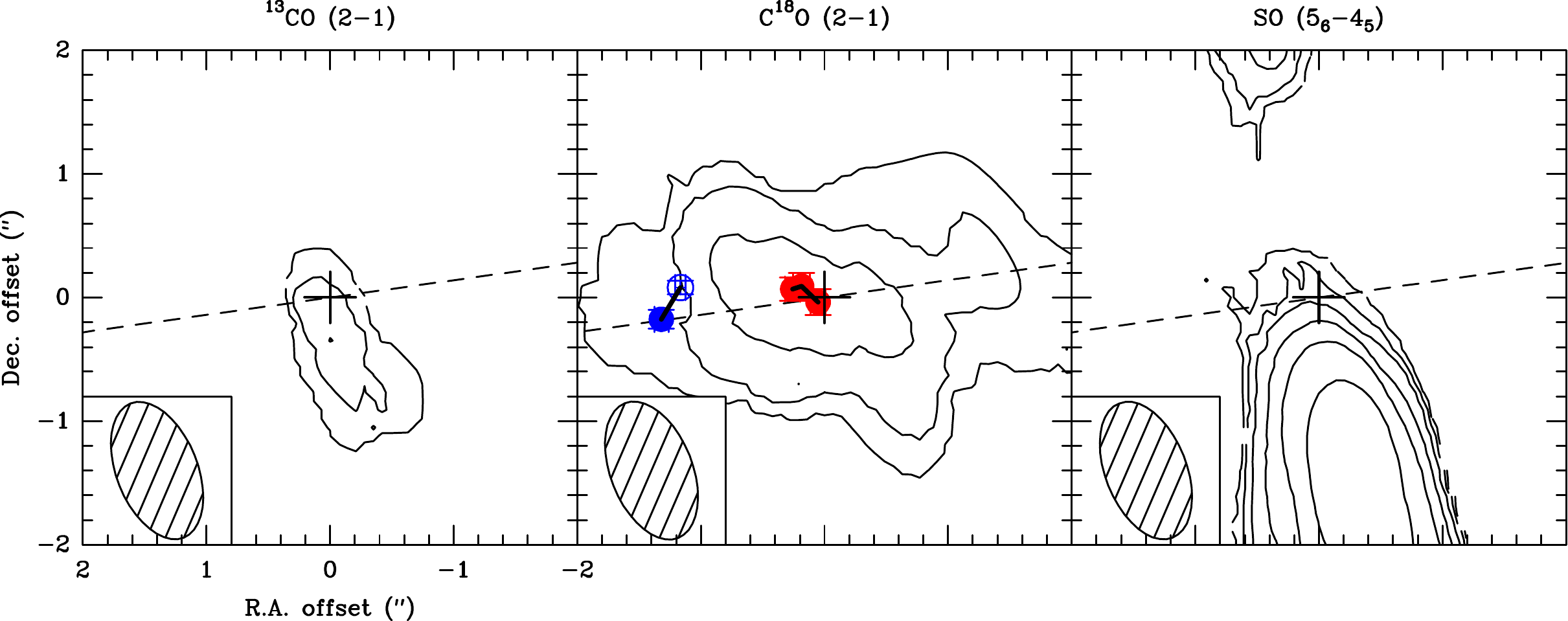}
  \caption{Same as in Fig.~\ref{fig:uvfit-l1527} for
    SerpS-MM18. \label{fig:uvfit-serps-mm18}}
\end{figure*}

\begin{figure*}
  \includegraphics[width=\hsize]{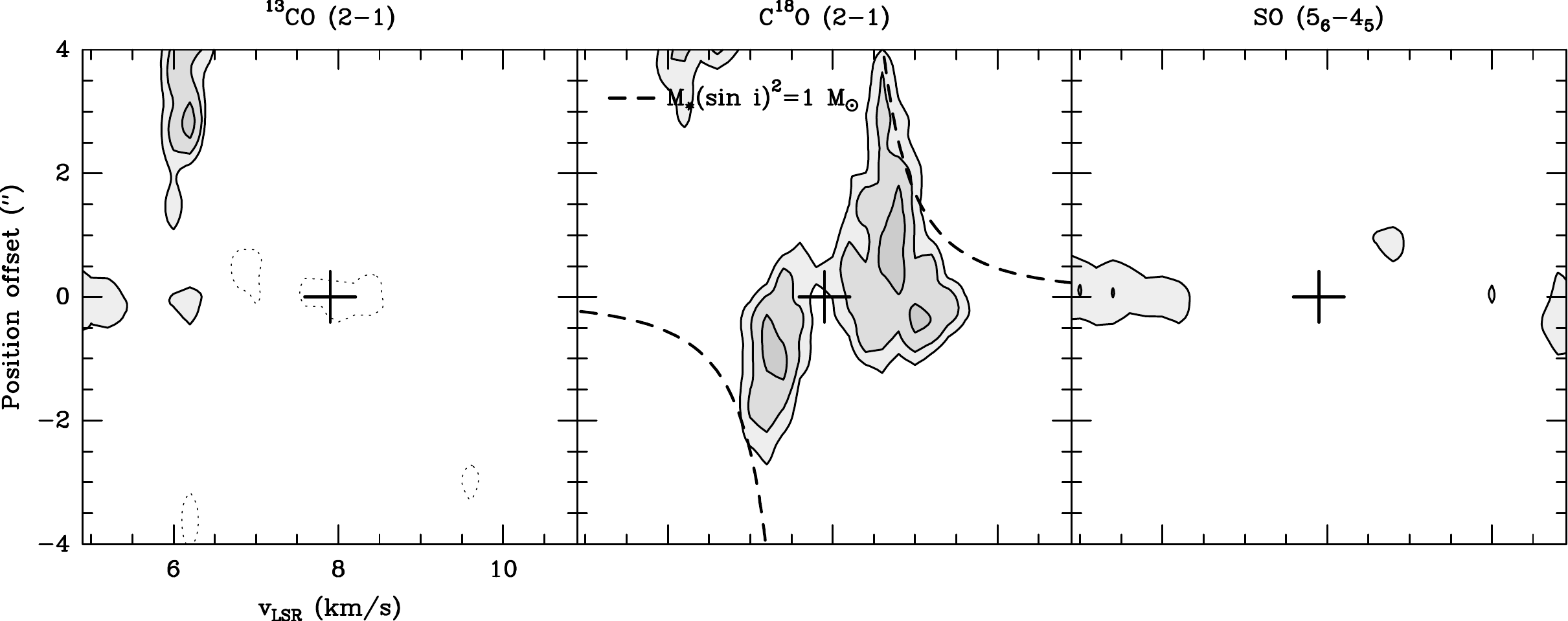}
  \caption{Same as in Fig.~\ref{fig:pv-l1527} for SerpS-MM18. The
    dashed curve shows a Keplerian velocity profile for
    $\Msinisqr = 1.0~\Msun{}$. \label{fig:pv-serps-mm18}}
\end{figure*}

Figure~\ref{fig:uvfit-serps-mm18} shows the results of the fit of
centroids of the emission in each channel, together with the assumed
disk major axis (PA -42\degr{}, orthogonal to the jet axis). No
centroids can be fitted for the \tcoto{} and \sofs{} lines. For the
\ceoto{} line, we fit the centroids in 5 channels. Surprisingly, the
centroids for the blue shifted and red-shifted channels are all
located east of the continuum peak\footnote{In principle this could be
  explained if the systemic velocity we adopted for the source
  (7.9~\kms{}) was too low. However, the PV diagram for the \ceoto{}
  line is consistent with the adopted \vsys{}.}. Because of this, we
cannot fit a velocity curve for this source. The PV diagrams along the
assumed disk axis are shown on Fig.~\ref{fig:pv-serps-mm18}. The PV
diagram for the \ceoto{} emission is consistent with rotation and
perhaps infall. Fitting the first contour with a Keplerian law gives
$\Msinisqr = 0.6~\Msun{}$.

To summarize the results for this source, the \ceoto{} PV diagram
indicates the rotation of the envelope in the direction perpendicular
to the jet. Because of outflow emission, we cannot fit a velocity
profile and we therefore have no evidence for Keplerian rotation in
this source.

  \section{Test of the rotation curve determination from the visibilities}
\label{sec:benchm-rotat-curve}

In this section, we test the technique we use in
Sect.~\ref{sec:disk-candidates} to derive the rotation curves for the
disk candidates from the visibilities. For this, we use the {\tt
  thindisk} code \citep{sebastien_maret_2019_3244265} to compute the
line emission of a geometrically thin Keplerian disk. The disk line
surface brightness distribution is assumed to follow a
tapered power-law distribution \citep{2009ApJ...700.1502A}:

\begin{equation}
  I_\mathrm{peak} \left( r \right ) \propto \left( \frac{r}{r_\mathrm{c}}
  \right)^{-\gamma} \, \mathrm{exp} \left( - \left( \frac{r}{r_\mathrm{c}}
    \right)^{2-\gamma} \right)
  \label{eq:d1},
\end{equation}
  
\noindent
where $r$ is the radius, $r_{c}$ is the disk characteristic
radius\footnote{\rev{The characteristic radius (defined in
    Eq.~\ref{eq:d1}) does not mean the disk-envelope boundary, because
    our model assumes Keplerian rotation outside the characteristic
    radius.}}, and $\gamma$ is the power-law exponent. We assume that
$r_{c} = 100$~au and $\gamma = 0.8$ \citep{2013A&A...557A.133D}. We
also assume that the disk inclination is 45\degr, and that the disk
major axis PA is 0\degr{}. The disk is assumed to be in Keplerian
rotation around a central mass $\Mstar = 0.3$~\Msun{}. Finally, we
adopt a distance of 293~pc (the same distance as L1448-C) and a
systemic velocity \rev{of} $v_\mathrm{sys} = 5$~\kms{}. \rev{The model
  does not include artificial noise.}

The {\tt thindisk} code produces a synthetic data cube that we process
with {\tt GILDAS} to simulate the observations with the
interferometer. First we use the {\tt UV\_FMODEL} task to compute
synthetic visibilities, assuming the same $uv$ coverage \rev{as} our L1448-C
observations. Then the visibilities are imaged using robust weighting
and deconvolved. Finally, we analyze the synthetic observations using
the technique outlined in Sect.~\ref{sec:l1527}.

\begin{figure*}
  \includegraphics[width=\hsize]{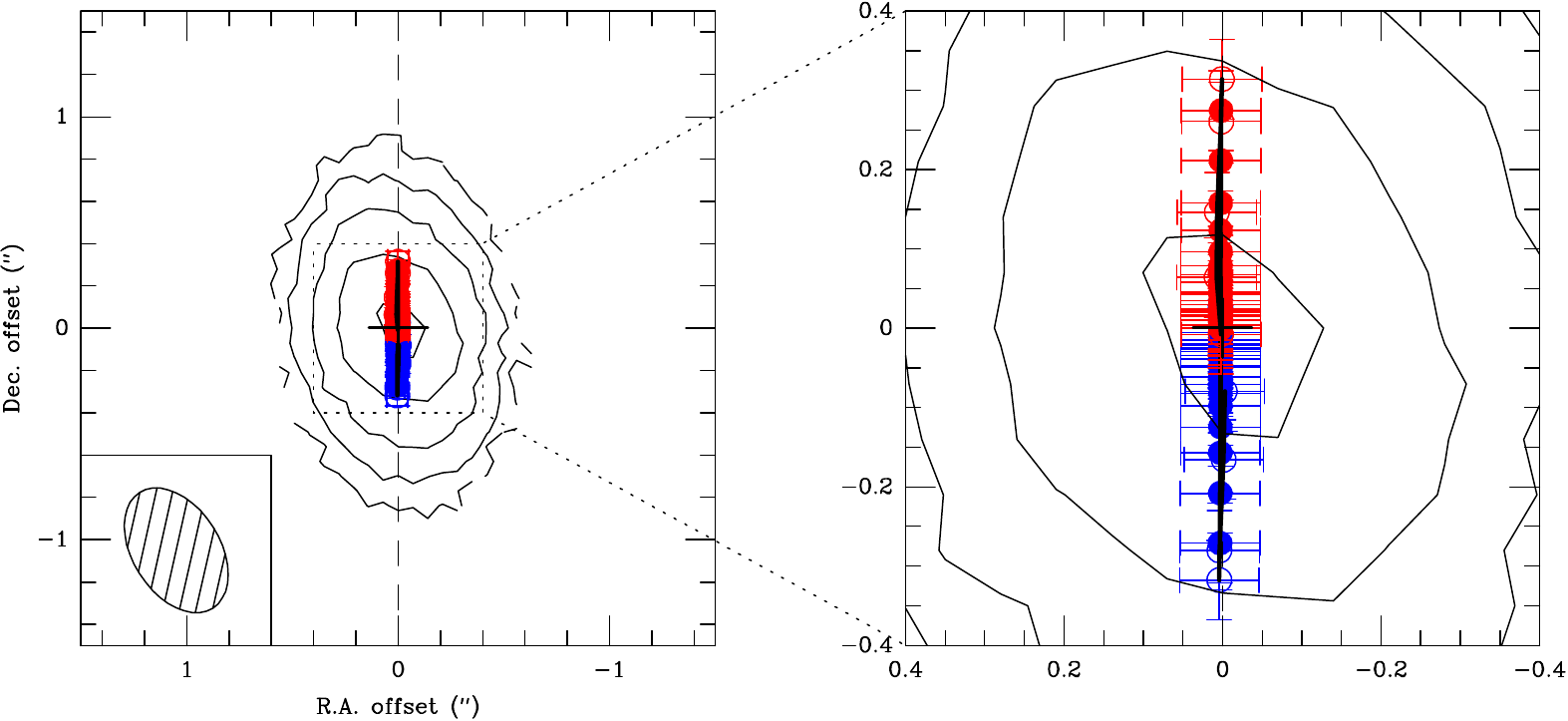}
  \caption{Same as in Fig.~\ref{fig:uvfit-l1527} for the synthetic
    disk model. The model assumes a disk characteristic radius
    $r_{c} = 100$~au, a disk inclination of 45\degr, a disk major axis
    PA of 0\degr{}, a central mass $\Mstar = 0.3$~\Msun{} and a
    distance of 293~pc.  The right panel is a zoom on the central
    region of the plot in the left panel.\label{fig:uvfit-thindisk}}
\end{figure*}

\begin{figure}

\end{figure}

Figure~\ref{fig:uvfit-thindisk} shows the results of the fit of the
centroid position in the $uv$ plane for each spectral channel
superimposed on the zeroth-order moment contours for the model. In
this figure, we see that the centroid positions are located onto the
disk major axis, with the blue-shifted and red-shifted channels
located \rev{in the} south and north of the disk central position,
respectively. The zeroth-order moment contours have an ellipsoidal
shape with a major axis orientation that appears to be slightly tilted
with respect to the north-south axis. This is an effect of the
synthetic beam, whose FWHM size is $0.65\arcsec \times 0.41\arcsec$
and PA is 32\degr{}. The fit in the $uv$ plane is obviously
insensitive to the synthetic beam orientation and shape, which
explains why the centroid positions in each channel are well aligned
with the true disk major axis.

\begin{figure}
  \includegraphics[width=\hsize]{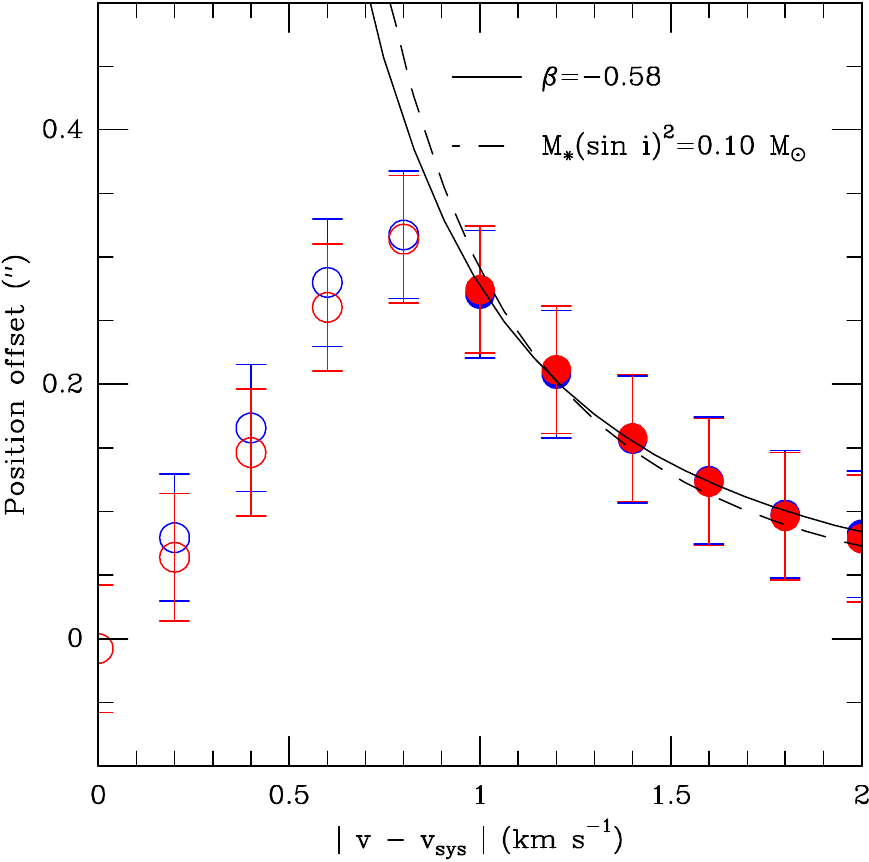}
  \caption{Same as in Fig.~\ref{fig:velocityfit-l1527} for the
    synthetic disk model. \label{fig:velocity-thindisk}}
\end{figure}

\begin{figure}
  \includegraphics[width=\hsize]{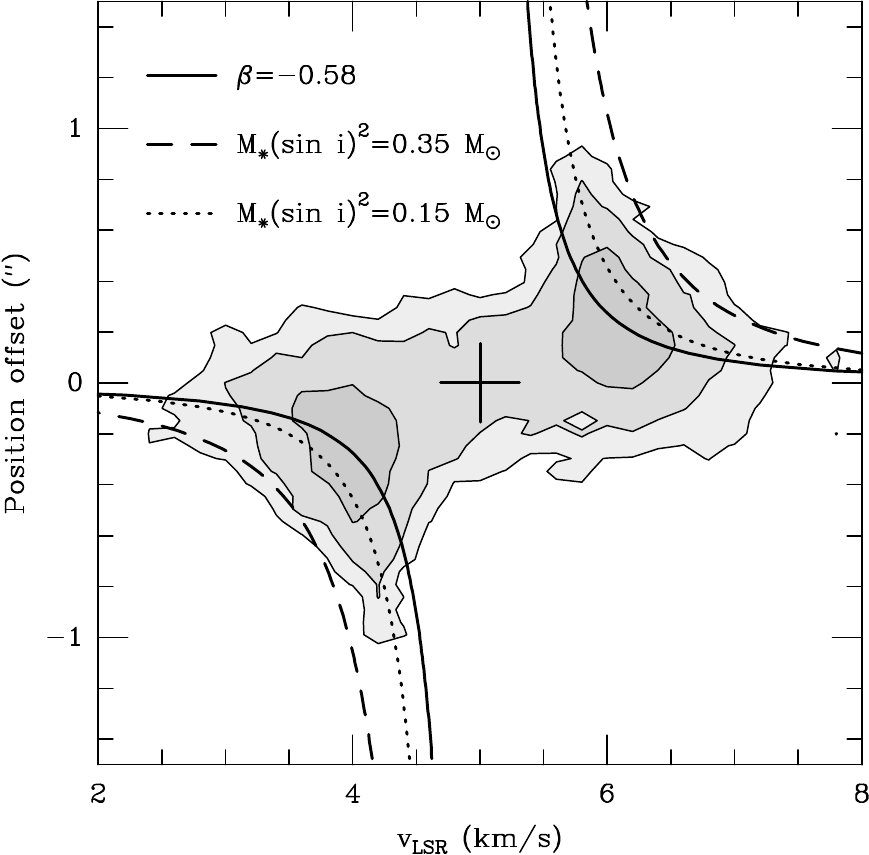}
  \caption{PV diagram along the disk major axis for the synthetic disk
    model. The solid line shows the velocity curve obtained from a fit
    in the $uv$ plane \rev{(see the solid curve in
      Fig.~\ref{fig:velocity-thindisk})}. The dashed line shows a fit
    of the first emission contour with a Keplerian law for
    $\Msinisqr{} = 0.35$~\Msun{}, that is\rev{,}
    $\Mstar{} = 0.7$~\Msun{}. \rev{The dotted line shows the Keplerian
      law for the underlying disk model, with
      $\Msinisqr = 0.15$~\Msun{}, that is, $\Mstar = 0.3
      $~\Msun{}.} \label{fig:pv-thindisk}}
\end{figure}

Figure~\ref{fig:velocity-thindisk} shows the offset along the disk
major axis for each velocity channel, as a function of the velocity
offset. This plot shows a similar pattern than the L1527 plot for SO
(see Fig.~\ref{fig:velocityfit-l1527}): a linear increase\footnote{We
  note that because of the finite spatial resolution, the model
  predicts a linear variation although the \rev{velocity} profile is a
  power law. This demonstrates that a linear variation in the rotation
  curve does not necessarily mean that the line arises from a ring in
  the disk \citep[see\rev{,} e.g.\rev{,}][]{Ohashi14}. This conclusion is not
  specific to the $uv$ plane fitting technique: a fit of peak emission
  as a function of velocity in the PV diagram shown in
  Fig.~\ref{fig:pv-thindisk} would also give a linear variation for
  velocities between $~$4 and $~$6 \kms{}.}  position offset as a
function of the velocity offset between
$0 < | v - v_\mathrm{sys} | \le 0.8$~\kms{}, and a power-law decrease
at higher velocity offset. Fitting the points at
$| v - v_\mathrm{sys} | > 0.8$~\kms{} with a power law, we find a
best-fit $\beta = -0.58 \pm 0.13$, consistent with a Keplerian
profile. With a Keplerian law, we obtain a best fit
$\Msinisqr{} = 0.10 \pm 0.01$~\Msun{}, that is,
$\Mstar{} = 0.20 \pm 0.02$~\Msun{} after correcting for the
inclination. This is about 30\% lower than the model central
mass\footnote{\rev{This correction factor is larger than the one
    derived analytically by \citet[23\%;][]{Aso15}}.}. We can estimate
the size of the disk from the centroid with the largest position
offset that is consistent with the Keplerian velocity curve, within
the error bars. On Fig.~\ref{fig:velocity-thindisk} we see that
Keplerian rotation is detected up to position offset of
$\sim 0.3\arcsec$. This gives a disk radius of 88~au for the assumed
distance. This radius is comparable to the characteristic radius
adopted in our model.

Figure~\ref{fig:pv-thindisk} shows the PV diagram along the disk major
axis, together with the best-fit rotation curve. Because the rotation
curve is based on fits of the centroid position in each velocity
channel, it follows the position of peak intensity in the PV
diagram. Therefore, it tends to underestimate the velocity of the disk
at each position, and in turn the mass of the central object. The mass
of the disk can also be estimated by a fit of the first emission
contour in the PV diagram. Using this technique, we find
$\Mstar{} = 0.7$~\Msun{}. This is about a factor of two larger than
the assumed central mass\footnote{This \rev{is} an effect of the
  spatial resolution: because the disk is marginally resolved, the
  emission contours are broaden by the synthetic beam along the
  position offset axis. This makes the first contour to appear further
  away from the central object than it would if the disk was resolved
  spatially. In turn, this tends to overestimate the mass.}  To
summarize, we show that our rotation curve technique provides a good
estimate of the power-law index of the rotation curve, even for a
marginally spatially resolved disk. However, it underestimates the mass
of the central object by about 30\% for the adopted model. On the
other hand, a fit of the first contour in the PV diagram overestimates
the mass of the central object by about a factor of two. By combining
these two techniques, we can obtain lower and upper limits on the mass
of the central object. In addition, our technique can be used to
estimate the disk's characteristic radius.

\end{appendix}

\end{document}